\documentclass[12pt]{article}
\usepackage{amsmath,amssymb,graphicx,mathrsfs,hyperref,slashed,abstract}

\newcommand{\be}{\begin{equation}}
\newcommand{\ee}{\end{equation}}
\newcommand{\bea}{\begin{eqnarray}}
\newcommand{\eea}{\end{eqnarray}}

\newcommand{\wh}{\widehat}
\newcommand{\wt}{\widetilde}

\newcommand{\on}{\overline N}

\def\({\left(} \def\){\right)}
\def\N{{\cal N}}
\renewcommand{\baselinestretch}{1.2}\normalsize
\begin{document}

\title{\vspace{-1.8in}
\begin{flushright} {\footnotesize LMU-ASC 71/13 }  \end{flushright}
{Phases of information release \\ during black hole evaporation}}
\author{\large Ram Brustein${}^{(1,2)}$,  A.J.M. Medved${}^{(3)}$ \\
\vspace{-.5in} \hspace{-1.5in} \vbox{
 \begin{flushleft}
  $^{\textrm{\normalsize
(1)\ Department of Physics, Ben-Gurion University,
    Beer-Sheva 84105, Israel}}$
  $^{\textrm{\normalsize  (2) CAS, Ludwig-Maximilians-Universit\"at M\"unchen, 80333 M\"unchen, Germany}}$
$^{\textrm{\normalsize (3)  Department of Physics \& Electronics, Rhodes University,
  Grahamstown 6140, South Africa }}$
 \\ \small \hspace{1.7in}
    ramyb@bgu.ac.il,\  j.medved@ru.ac.za
\end{flushleft}
}}
\date{}
\maketitle
\begin{abstract}

In a recent article, we have shown how quantum fluctuations of the background geometry modify Hawking's density matrix for black hole (BH) radiation. Hawking's diagonal matrix picks up small off-diagonal elements whose influence becomes larger with the number of emitted particles.  We have calculated the ``time-of-first-bit", when the first bit of information comes out of the BH, and the ``transparency time", when the rate of information release  becomes order unity. We have found that the transparency time is equal to  the ``Page time'', when the BH has lost half of its initial entropy to the radiation, in agreement with Page's results. Here, we improve our previous calculation by keeping track of the time of emission of the Hawking particles and their back-reaction on the BH. Our analysis reveals a new time scale, the radiation ``coherence time'', which is  equal to the geometric mean of the evaporation time and the light crossing time.  We find, as for our previous treatment, that the time-of-first-bit is equal to the  coherence time, which is much shorter than the Page time. But the  transparency time is now much later than the Page time, just one coherence time before the end of evaporation. Close to the end, when the BH is parametrically of Planckian dimensions but still large, the coherence time becomes parametrically equal to the evaporation time, thus allowing the radiation to purify. We also determine the time dependence of the entanglement entropy of the early and late-emitted radiation. This entropy is small during most of the lifetime of the BH, but our qualitative analysis suggests that it becomes parametrically maximal near the end of evaporation.
\end{abstract}

\newpage
\renewcommand{\baselinestretch}{1.5}\normalsize

\section{Introduction}

That  black holes (BHs) radiate thermally was a remarkable finding \cite{Bek,Hawk} but  has also lead to some infamous puzzles.
For instance, an initially pure state of matter can collapse to form a BH and eventually evaporate into a mixed state of thermal radiation. This is in direct conflict with quantum mechanics, which postulates a unitary time evolution and so forbids a pure state from evolving into a mixed one.
This is, in essence,  the BH information-loss paradox \cite{info}. (For reviews, see \cite{info2,info3,info4}.)

Over the years, a myriad of explanations has been suggested on how this tenuous situation gets resolved. Initially, Hawking thought that the laws of quantum mechanics have to be changed. Others have  sometimes claimed that a theory  more fundamental than general relativity, such as string theory, or some exotic new physics, such as highly entropic remnants,  is needed to resolve the matter. However, strong circumstantial evidence has been gathered, indicating that general relativity and ordinary quantum mechanics are sufficient for consistently describing the process of BH evaporation. In particular, the quantum-information treatments of Page \cite{page} and then of Hayden and Preskill \cite{HaydenPreskill} demonstrate that a slowly burning matter system  ---  be it the complete works of Shakespeare or a Schwarzschild BH --- must emit all of its information by the end of evaporation. Consequently, a thermal mixed state  cannot be the final product. Once this logic is accepted, the challenge then becomes to identify what is still  missing from the standard treatments; namely, the  information-release mechanism that is responsible for restoring unitarity by the end of the BH evaporation. The review articles \cite{info2,info3,info4} contain further discussion of the issues concerning the BH information paradox.

In \cite{RB}, it was proposed that the origin of the BH information paradox is the use of a strictly classical geometry for the BH. (See \cite{Dvali1,Dvali2,Dvali3} for overlapping ideas.)  It was also argued  that the leading semiclassical corrections that account for the quantum  fluctuations of the background geometry should be taken into account by assigning a wavefunction to the BH.  The contention was that the parameter which controls the strength of the semiclassical corrections is the ratio of the Compton wavelength of the BH $\;\lambda_{BH}=\hbar/M_{BH}\;$ to its radial size $R_S$. In \cite{flucyou},  we have proposed a concrete scheme for evaluating the semiclassical corrections using the wavefunction of \cite{RM,RB}. The parameter that controls the strength of the semiclassical corrections was denoted by $C_{BH}$ and  determined more precisely,
$\;C_{BH}=1/S_{BH}=\frac{\lambda_{BH}}{2\pi}/R_S\;$.

We have, in a recent article \cite{slowleak},   gone on to apply this idea to the calculation of the Hawking radiation.
There, Hawking's calculation was repeated but with one additional input: The assignment of a  Gaussian  wavefunction to the collapsing shell of  matter. The main distinction between our treatment and Hawking's  is the introduction of a new scale, the width of the wavefunction. On the basis of the Bohr correspondence principle \cite{RM,RB}, this width should be Planckian.

 After computing the appropriate expectation values, we obtained a picture that is different than that found by Hawking and consistent with Page's. Most pertinently, Hawking's density matrix for the BH radiation is strictly diagonal whereas our matrix contains small off-diagonal elements of order $\sqrt{\hbar}$  in the same basis. The effect of these elements on the eigenvalues of the matrix is initially small  but, as the number $N$  of emitted particles grows, so does the changes to the eigenvalues. The parameter that controls these changes to the matrix was found to be $N C_{BH}$.

We have calculated the time when the rate of information release  becomes of  order unity. This  ``transparency time'' $t_{trans}$  was found to coincide with the time when  $\;N C_{BH}=1\;$, which  is, in turn, the same as the ``Page time''  when the BH has lost half of its initial entropy to the radiation. Hence, this result is  in agreement  with  the analysis  of  Page \cite{page}.

We have also calculated  the ``time-of-first-bit" $t_{1bit}$, when the first bit of information comes out of the BH. This occurs when $\;N^2C_{BH}=1\;$, which is much earlier than the  Page time and apparently in disagreement with Page's calculation. However, we have tracked the information in  the correlations between the shell and radiation as well as  in the radiation subsystem.  Page, on the other hand, tracked only the latter, which is an exponentially suppressed quantity until after the transparency time.

To keep the  calculations in \cite{slowleak} as simple as possible, we have ignored the fact that the Hawking particles are emitted over a time scale spanning the lifetime of the BH. In effect, we were assuming that all the Hawking particles are  being emitted coherently. Here, we will improve upon our previous calculation by keeping track of  the time of emission of the Hawking particles and their back-reaction on the BH.  Our analysis reveals a new time scale --- the radiation coherence time $\;t_{coh}=R_S^2/l_p\;$  ---
which is  the geometric mean of the evaporation time  $R_S^3/l_p^2$ and the light crossing time $R_S$.
The number of coherent Hawking particles $N_{coh}$ that is  emitted during this time is typically  given by $\;N_{coh}=1/\sqrt{C_{BH}}=\sqrt{S_{BH}}\;$.
This estimate for $N_{coh}$ is valid during most of the lifetime of the BH
but gets modified at the last stages of evaporation  (see below).

What we   find is  that the Page time splits into two: The time-of-first-bit is the same as found before: much earlier than the Page time.  It can now be  identified with the coherence time,    $\;t_{1bit}=t_{coh}\;$.  On the other hand, the transparency time turns
out to be much  later than the Page time. This phase  now  transpires at  one coherence time before the time of final evaporation, $\;t_{final}-t_{trans}=t_{coh}\;$,  which happens when $\;N_{coh} C_{BH} =1\;$, so that
$N_{coh} C_{BH}$ now  replaces  $NC_{BH}$  as the parameter controlling  the corrections  to Hawking's diagonal density  matrix.

Let us explain the origin of the main difference between our picture of BH evaporation and Hawking's. The advanced time of particle emission  and the frequency of the emitted particle are conjugate variables. In our description, this translates (as in Eq.~(\ref{tdnotnine}) of  the Appendix) into the following  conjugate pair: the dimensionless  frequency $\omega$  and $C_{BH}\Delta N$,  where $\Delta N$ is the  number of  particle emissions that have occurred since the time of  emission of some specific particle. The width in  $C_{BH}\Delta N$ is therefore determined by the inverse of the width in  $\omega$.  In Hawking's calculation, the canonical
relation between $\omega$ and $\Delta N$ does not exist because $C_{BH}$  vanishes. Consequently, the widths in both $\omega$ and $\Delta N$ vanish ({\em i.e.}, they are controlled by delta functions). However, for our Gaussian wavefunction,  the width in $\omega$ is proportional to $C_{BH}^{-1/2}$ ({\em cf}, Eq.~(\ref{j6})).  For $\Delta N$, the width is determined by the inverse of
the width in  $\omega$, and so it is proportional to $\;C_{BH}^{1/2}/C_{BH}= C_{BH}^{-1/2}\;$.  The width in $\Delta N$ then determines the coherence scale, $\;N_{coh}\sim C_{BH}^{-1/2}\;$.

The situation changes at the late stages of evaporation, although
this era can only be discussed in a qualitative way because our methods
become less accurate for this region of parameters.
Here, the BH is parametrically nearing Planckian dimensions, but still large and semiclassical, so  that $\;C_{BH}$ is becoming larger and approaching order unity. It is  clear that  the width of $\omega$ is decreasing and approaching unity but, somewhat  surprisingly, the width of  $\Delta N$  is  growing. To understand this unexpected result, note that the factor $C_{BH}$ in the product $C_{BH}\Delta N$ is determined by the time of emission of the Hawking particle and is small for most emitted particles. Therefore,  $N_{coh}$  goes at the end as $\;1/C_{BH}\sim S_{BH}\;$;  where the  $S_{BH}$ means the BH entropy at an earlier epoch, so that $\;N_{coh}\gg 1\;$. Based on this qualitative analysis, it will be argued that, by this point in the evaporation, the entirety of the emitted Hawking particles become coherent,  $\;N_{coh}\sim$ total number of particles. Consequently,  the radiation purifies at a high rate.

The  sequence by which the correlations between the emitted particles evolves now  becomes  clear: Always a delta function for Hawking since $N_{coh}=1$. While in our case, first, a smoothed delta function when $\;N_{coh} \sim 1/\sqrt{C_{BH}}\;$,  followed by a theta function when $\;N_{coh}\sim 1/C_{BH}\;$.

Taking into account the time-dependent emissions, we  are able to  determine  the evolution of the entanglement between the early and late radiation. We find that this entanglement is initially very small but   becomes significant  at  $t_{trans}$ and  then grows  quickly to be (parametrically) maximal when the radiation purifies.

The rest of the paper proceeds as follows: First, in Section~2, we summarize our preceding work \cite{slowleak}. This is essential for understanding the remainder, as we adopt  this initial framework and build up the  analysis from there.
Next, in Section~3, we determine what is the effect of time-dependent emissions  and use this to better understand how the evaporation process evolves. In Section~4,  a calculation of the trace of the square of the density matrix enables us  to analyze the rate of information transfer, to quantify  the various time scales and to qualitatively demonstrate  that  the radiation  does indeed become purified by the end. Then, in Section~5, we calculate the entanglement entropy for the early- and late-emitted radiation and qualitatively show that it becomes parametrically maximal at late times. The paper concludes with a summary (Section~6) and an appendix.

Recently,  a modern interpretation of the information-loss paradox, known
as the ``firewall'' problem \cite{AMPS} (also see \cite{Sunny1,info4,Braun} for earlier versions and \cite{fw1,fw2,fw3,fw4,Sunny2,avery,lowe,vv,pap,AMPSS,SMfw,pagefw,VRfw,MP} for a sample of the related literature). We expect that the current analysis will be an essential step toward a resolution of this puzzle, but defer  a specific discussion to a future publication \cite{inprog}.

\section{Review of previous results on semiclassical corrections to Hawking radiation}

\subsection{Conventions}

We will now review our preceding paper \cite{slowleak}, which the reader can
refer to for an in-depth discussion. This review will also serve to introduce notations and conventions that we will use in the following sections.

We choose units such that Planck's constant $\hbar$ and Newton's constant $G$ are explicit, and all other fundamental constants are set to unity. In some instances,
the Planck length $l_p=\sqrt{\hbar G}$ is used instead.

We assume a four-dimensional Schwarzschild BH  (generalizations to higher dimensions are straightforward) of large but finite mass $\;M_{BH}\gg\sqrt{\hbar/G}\;$, with the  metric $\; ds^2 = -\left(1-\frac{R_{S}}{r}\right)dt^2+ \left(1-\frac{R_{S}}{r}\right)^{-1}dr^2 +d\Omega^2_2\;$. Here, $\;R_{S}=2M_{BH}G\;$  denotes the horizon radius.

We use a  dimensionless advanced-time coordinate $\;v =\frac{1}{R_S}\left(t+r^{\ast}\right)\;$, where $\; r^{\ast}=\int^r\;dr \sqrt{-g^{tt}g_{rr}} = r+R_{S}\ln(r-R_{S})\;$. Thus, our  frequencies are also dimensionless in units of inverse Schwarzschild radius.

For a Schwarzschild  BH, the values of its Hawking temperature and Bekenstein--Hawking entropy are $\;T_{H}=\frac{\hbar}{4\pi R_{S}}\;$
and  $\;S_{BH}=\frac{\pi R_{S}^2}{\hbar G}\;$.

A BH is often meant as shorthand  for ``a  collapsing shell of matter that goes
on to form a BH''. Technically, in our calculations, all particles are emitted before
the horizon actually forms, as in Hawking's original calculation.

\subsection{Semiclassical density matrix}

The objective  of \cite{slowleak} was to calculate  the modifications due to a fluctuating geometry to Hawking's thermal density matrix for the radiation emitted by a collapsing shell of matter. As the geometry is sourced by the collapsing shell,   we have assigned it a wavefunction,
\be
\left.\Psi_{shell}(R_{shell})\right|_{R_{shell}\to R_{S}}\; = \;{\cal N}^{-1/2} e^{-\frac{\left({R}_{shell}-R_{S}\right)^2}{2C_{BH} R_{S}^2}}
\label{wave}
\;,
\ee
where  $R_S$ is the Schwarzschild radius of the incipient BH,  $R_{shell}$ is the radius of the shell, $\;{\cal N}\simeq 4\pi R_{S}^2\sqrt{\pi\hbar G}\;$ is a normalization constant and $\;C_{BH}=S_{BH}^{-1}\;$ is
the aforementioned ``classicality'' parameter. This form of BH wavefunction was first justified in \cite{RM} and then further motivated in \cite{RB,flucyou,slowleak}.

The classicality parameter $C_{BH}$ can be viewed as a dimensionless (scaled) $\hbar$ that evolves in time, $\;C_{BH}=\hbar(t)\;$.  It is initially  very small  for a large BH but steadily grows as the BH evaporates.  In this sense,  semiclassical corrections to observables can be expressed as powers of this dimensionless $\hbar$.

For a discussion of BH radiation, it is more convenient to use the advanced time of the shell $v_{shell}$. The conversion to $\Psi_{shell}(v_{shell})$ is made by observing that, in the near-horizon limit,  $\;\frac{R_{shell}-R_S}{R_S} \simeq \left(v_0-v_{shell}\right)\;$, where $v_0$ is the advanced time at which the shell crosses its horizon.  One can then compute the expectation value of a generic operator $O$ as follows:
\bea
 \langle  \wh{O}(R_{shell})\rangle\;=\;
\frac{4\pi}{\N} \int\limits_0^\infty dR_{shell}\; R_{shell}^2\ e^{-\frac{(R_{shell}-R_S)^2}{2 \sigma^2}} {O}(R_{shell}) \hspace{1in}
\cr
\;\simeq\; \frac{4\pi R_S}{\N}  \int\limits_{-\infty}^{\infty} dv_{shell}\left[ R_S^2 + 2R^2_S(v_0-v_{shell})+R_S^2(v_0-v_{shell})\right]\ e^{-\frac{(v_0-v_{shell})^2}{C_{BH}}}{O}(v_{shell})\;,
\label{vfunction}
\eea
where $\;\sigma^2= R_S^2C_{BH}/2= l_p^2/2\pi\;$.
Equation~(\ref{vfunction}) is a particular case  of our more general prescription \cite{RB,flucyou}, which amounts to applying the standard rules of quantum mechanics.

Hawking's calculation \cite{Hawk} relates the in-going modes with the out-going modes (the Hawking particles) by a  Bogolubov transformation,
\be
F_{\omega} \;=\; \int\limits_{0}^{\infty} \;d\omega^{\prime}\;\Big(\alpha_{\omega^{\prime}\omega}f_{\omega^{\prime}}
+\beta_{\omega^{\prime}\omega}f^{\ast}_{\omega^{\prime}}\Big)\;.
\label{prestuff}
\ee
Here,  $F_{\omega}$
is an out-mode that has been traced back to past null infinity,
$\;f_{\omega^{\prime}}=\frac{1}{\sqrt{2\pi}}e^{i\omega^{\prime}v}\;$
is a basis function for an in-mode and the $\alpha$'s ($\beta$'s) are
the positive-energy (negative-energy) Bogolubov coefficients. Recall that, unlike Hawking (and unlike in \cite{slowleak}),
we are using dimensionless frequencies.

The Hawking single-particle density matrix for the out-modes can then be expressed as
\bea
\rho_H(\omega,\wt{\omega}) &=&
\langle 0_{in}| F_{\omega}^{\ast}F_{\widetilde{\omega}}|0_{in} \rangle\; \nonumber \\
&=&\int\limits_{-\infty}^{v_0} dv \int\limits_{0}^{\infty} d\omega^{\prime} \int\limits_{0}^{\infty} d\omega^{\prime\prime}
\;\beta^{\ast}_{\omega^{\prime}\omega}\beta_{\omega^{\prime\prime}\widetilde{\omega}}
 \frac{e^{iv\left(\omega^{\prime}-\omega^{\prime\prime}\right)}}{2\pi}\;,
\label{rhoH}
\eea
with $|0_{in}\rangle$ denoting the vacuum
annihilated by positive-frequency in-modes.

Hawking used a procedure of ray tracing that exploited the geometric optics of the modes to determine that
\bea
\;\beta_{\omega^{\prime}\omega}&\propto& \frac{1}{2\pi}\int\limits_{-\infty}^{v_0}dv\ e^{i \omega' v}\ e^{-i 2 \omega  \ln(v_0-v)}
\cr &=& \Gamma\left(1-i2\omega \right) \left(i\omega^{\prime}\right)^{-1+i2
\omega}
\frac{1}{2\pi} e^{iv_0 \omega^{\prime}}\;.
\label{betaH}
\eea
The logarithm in the top line takes into account the discontinuity in the phase of the modes as they pass across the shell at an advanced
time $v$ close to $v_0$. This phase discontinuity turns out to be central to our findings.

As discussed in \cite{slowleak}, only the Bogolubov coefficients are sensitive to the effects of the fluctuating background geometry. Hence, applying our prescription~(\ref{vfunction}), we obtain the ``semiclassical'' density matrix,
\be
\rho_{SC}(\omega,\wt{\omega})
\;=\;\int\limits_{-\infty}^{v_0} dv \int\limits_{0}^{\infty} d\omega^{\prime} \int\limits_{0}^{\infty} d\omega^{\prime\prime}
\langle\Psi_{shell}|\;\beta^{\ast}_{\omega^{\prime}\omega,~SC}~ \beta_{\omega^{\prime\prime}\widetilde{\omega},~SC}|\Psi_{shell}\rangle \frac{e^{iv\left(\omega^{\prime}-\omega^{\prime\prime}\right)}}{2\pi}\;.
\label{rhoSCD}
\ee
The ``semiclassical'' coefficients $\beta_{\omega^{\prime\prime}\widetilde{\omega},~SC}$ are derived in the same way as Hawking does but now take into account that the discontinuity in the phase depends on $v_{shell}$ rather than on  $v_0$,
\bea
\beta_{\omega^{\prime}\omega,~SC}&\propto& \frac{1}{2\pi} \int\limits_{-\infty}^{v_{shell}}dv\ e^{i \omega' v}\ e^{-i 2 \omega  \ln(v_{shell}-v)}
\cr &=& \Gamma\left(1-i2\omega \right) \left(i\omega^{\prime}\right)^{-1+i2
\omega }
\frac{1}{2\pi} e^{iv_{shell}\omega^{\prime}}\;.
\label{betaQ}
\eea

The $v$ integral in Eq.~(\ref{rhoSCD}) can be expressed as a sum of a classical term and the leading semiclassical correction. Denoting this integral as $\;{I}_{SC} =\frac{1}{2\pi}\int \limits_{-\infty}^{v_0} dv\;  e^{i (v-v_{shell}) (\omega'-\omega'')}\;$ and changing the  variable to  $\;v^{\prime}=v-v_{shell}\;$, we have
\bea
 {I}_{SC} &= & \frac{1}{2\pi}\int\limits_{-\infty}^{0} dv^{\prime}\;  e^{i v^{\prime} (\omega'-\omega'')}+\frac{1}{2\pi}\int
\limits_{0}^{v_0-v_{shell}} dv^{\prime}\;  e^{i v^{\prime} (\omega'-\omega'')}
\nonumber \\ &\equiv & I_C + \Delta I_{SC}(C_{BH})
\;.
\label{ISC}
\eea
The integral  on the left $I_C$ is a delta function $\delta(\omega-\omega^{\prime})$ and yields Hawking's classical result of a diagonal density matrix. The expectation value of the integral on the right  $\langle \wh{\Delta I}_{SC}(C_{BH})\rangle$  leads to the off-diagonal elements.

The expectation value of interest then goes as
\be
\langle \wh{\Delta I}_{SC}(C_{BH}) \rangle \;=\;\frac{4\pi R_S^3}{\N}\int\limits_{-\infty}^{\infty} d \wt{v}
\left[1 + 2\wt{v}+\wt{v}^2\right] e^{-\frac{\wt{v}^2}{C_{BH}}} \frac{1}{2\pi} \int\limits_{0}^{\wt{v}} dv^{\prime}\;  e^{i v^{\prime} (\omega'-\omega'')}\;,
\label{j2}
\ee
which was evaluated in \cite{slowleak} to leading order in $C_{BH}$,
\be
\langle \wh{\Delta I}_{SC}(C_{BH}) \rangle \;= \;\frac{1}{2\pi} C_{BH}\ e^{-\frac{(\omega'-\omega'')^2}{4} C_{BH}}\;.
\label{j6}
\ee

Substituting the full expressions for the Bogolubov coefficients \cite{Hawk} into Eq.~(\ref{rhoH}), we can write the semiclassical correction to Hawking's matrix as
 \bea
&&\Delta\rho_{SC}(\omega,\wt{\omega}) \;=\;
\frac{t^*_\omega t_{\wt{\omega}}}{(2\pi)^3} \frac{C_{BH}}{(\omega \wt{\omega})^{1/2}} \Gamma\left(1+i2\omega\right) \Gamma(1-i2\wt{\omega}) e^{-\pi(\omega+\wt{\omega})}   \cr &&
\;\;\;\;\;\times\;\int\limits_0^\infty d \omega'' \int\limits_0^\infty d \omega' \ e^{-\frac{(\omega'-\omega'')^2}{4} C_{BH}}\
(\omega')^{-1/2-i2\omega}
(\omega'')^{-1/2+i2\wt{\omega}}\;,\hspace{.5in}
\label{rhosc}
\eea
where $t_{\omega}$ is the transmission coefficient through the gravitational barrier. These remaining integrals can be computed analytically with some amount of effort. One caveat is a logarithmic divergence on the line $\;\omega=\omega^{\prime}\;$. We have handled this by isolating the divergent piece and then recognizing that this is just a small (order $C^{1/2}_{BH}$) correction to the diagonal matrix of Hawking.

The final result is  an off-diagonal correction of magnitude $C^{1/2}_{BH}$ to Hawking's classical matrix,
\bea
&&\Delta\rho_{SC}(\omega,\wt{\omega}~;C_{BH}) \;=\;
\frac{t^*_\omega t_{\wt{\omega}}}{(2\pi)^3} C_{BH}^{1/2} \frac{2}{(\omega \wt{\omega})^{1/2}}
\left(\frac{C_{BH}}{4}\right)^{+i2 (\omega-\wt{\omega})} \cr &\times&
 \Gamma\left(1+i2\omega\right)
\Gamma(1-i2\wt{\omega})
\ \hbox{\Large\em e}^{-\pi(\omega+\wt{\omega})} \
\Gamma\left(\frac{1}{2}- i(\omega-\wt{\omega})\right)\cr
&\times&\Biggl\{
\Gamma \left(i2 (\omega-\wt{\omega})\right)\left[
\frac{\Gamma \left(\frac{1}{2}+i2\wt{\omega}\right)}
{\Gamma \left(\frac{1}{2}+i2 \omega\right)}
+
\frac{\Gamma \left(\frac{1}{2}-i2\omega\right)}
{\Gamma \left(\frac{1}{2}-i2 \wt{\omega}\right)}
\right] + \frac{ i }{\omega-\wt{\omega}}\Biggr\}\;,
\label{rhoscf1}
\eea
which we  will subsequently denote as  $C_{BH}^{1/2}\Delta\rho_{OD}$ (for off-diagonal).

Recall that Hawking's classical matrix with dimensionless frequencies is given by
\bea
\rho_H(\omega, \widetilde{\omega})
 &=& \frac{t^*_\omega t_{\wt{\omega}}}{e^{4\pi\omega}-1} \delta(\omega-\widetilde{\omega})\;.
\label{spec}
 \eea
We will assume that the semiclassical matrix has  been renormalized to give  $\;{\rm Tr}\; \rho_H= \int d\omega\; \rho_H(\omega,\omega)=1\;$.

The next step in \cite{slowleak}  was constructing a multi-particle density matrix for $N$ identical, independent particles. This, in effect, amounts to the assumption that all the particles are coherent, so that the timing of their emissions does not affect the correlations among them. This will be corrected later.

This multi-particle  matrix consists of $N\times N$ blocks:  $\rho^{(N)}_{IJ}(\omega,\widetilde{\omega})$ with $\;I,J=1,\dots,N\;$ and any of the $N^2$ blocks  is a matrix of  the same dimensionality as the single-particle density matrix. Each diagonal entry is the single-particle Hawking matrix $\;\rho^{(N)}_{II}= \rho_H(\omega,\widetilde{\omega})\;$ (plus subdominant semiclassical corrections)  and each off-diagonal element contains the semiclassical part $\;\rho^{(N)}_{I\neq J}=C^{1/2}_{BH}\Delta\rho_{OD}(\omega,\widetilde{\omega})\;$. Each block is multiplied by a phase $e^{\Theta_{IJ}}$ ($\;\Theta_{IJ}=-\Theta_{JI}\;$), but these are of no consequence to our discussion.

The normalized $N$-particle density matrix can then be expressed as
\be
\rho^{(N)}_{IJ}(\omega,\widetilde{\omega}) \;=\; \frac{1}{N}
\rho_H(\omega,\widetilde{\omega})\mathbb{I}_{N\times N}
\;+\; \frac{1}{N}
C^{1/2}_{BH}\Delta\rho_{OD}(\omega,\widetilde{\omega}) \slashed{\mathbb{I}}_{N\times N} \;,
\label{nmatrix}
\ee
where the symbol $\slashed{\mathbb{I}}_{N\times N}$ denotes an $N\times N$ matrix of ones off the diagonal (up to the implied phases) and zeros on it.
This  matrix can be used to track the information flowing out of the BH.

\subsection{Entropy and information}

The von Neumann entropy per particle
$\;\frac{S}{N}=-{\rm Tr}\left[\rho^{(N)}\ln\rho^{(N)}\right]\;$ of the
radiation~\footnote{Alternatively, one can
symmetrize the particles and use the  normalization $1/N!\;$.
In which case, $\; S=-{\rm Tr}\left[\rho^{(N)}\ln\rho^{(N)}\right]\;$. The difference for large $N$ is insignificant.} can be calculated perturbatively in the small parameter $C_{BH}$.  This calculation yields, to leading order,
\be
S\;=\;
S_H\left(1- \frac{1}{2} K\ N C_{BH}\right)\;.
\label{staylor}
\ee
Here $S_H$ is the thermal entropy or, equivalently, the von Neumann entropy for the Hawking diagonal matrix. The coefficient $\;K=\frac{{\rm Tr}\left[ (\Delta\rho_{OD})^2 \rho_H^{-1} \right]}{-{\rm Tr}\left[ \rho_H \ln\rho_H\ \right]}\;$ is a positive numerical factor of order one.

From Eq.~(\ref{staylor}), it is possible to deduce that the parameter controlling the semiclassical corrections is $NC_{BH}$ rather than $C_{BH}$. This  outcome  is a consequence of having roughly $N$ times more off-diagonal elements than diagonal ones.  So that, when  $\;NC_{BH}=1\;$, the semiclassical corrections becomes large and one expects a significant change.

In \cite{slowleak},  the back-reaction of the Hawking particles on the geometry was incorporated in  the following (incomplete) way:  It was assumed  that the BH radiates thermally as a black body, which is clearly a good approximation during most of the lifetime of the BH. We further assumed that the radiated particles carry an energy equal to $T_H$, with the Hawking temperature $T_H$  also taken to be time dependent. Then  $\;dN=dM{\frac{dN}{dM}}=-\frac{dM}{T_H(t)}\;$, which can be integrated to give
\be
N(t)=S_{BH}(0)-S_{BH}(t)
\ee
and therefore, because $\;C_{BH}(t)=(S_{BH}(t))^{-1}\;$,
\be
\;C_{BH}(t)=\left[S_{BH}(0)- N(t)\right]^{-1}\;.
\ee
Also, since
\be
\;S_H(t)\simeq N(t)\;,
\label{shn}
\ee
it follows that $\;S_H(t)\simeq S_{BH}(0)-S_{BH}(t)\;$.

As $N(t)$ and $C_{BH}(t)$ are both monotonically increasing functions of time, their product is growing and will  eventually reach and then surpass unity.  Indeed, the transition out of the perturbative regime takes place at the  Page time \cite{page}, when the BH has lost half of its initial entropy to radiation. This finding appears to substantiate Page's claim that this moment represents a tipping point in the evaporation process.

We have also looked in \cite{slowleak} at the rate of information flow.
The information contained in the radiation is defined in the standard way,
\bea
I(t)\;=\; S_H(t)-S(t) \;&=&\; \frac{K}{2}S_H(t) N(t) C_{BH}(t) \cr &\simeq& \frac{K}{2}\frac{N(t)^2}{S_{BH}(0)-S_H(t)}
\;,
\label{inforef}
\eea
where Eq.~(\ref{shn}) has been applied (both here and below).

It  follows that
\be
\frac{dI}{dS_H}\;\simeq \;
\frac{K}{2}\left[2+C_{BH}N\right]
C_{BH}N\;,
\label{inforate}
\ee
and so $\frac{dI}{dS_H}$ is  small (order $C_{BH}$) before the Page time and of order unity at it. But, at later times, the previous  calculation  formally breaks down.

We can use Eqs.~(\ref{inforef}) and~(\ref{inforate}) to calculate $t_{1bit}$ and $t_{trans}$. Recall that $t_{1bit}$ is defined to be the time when the first bit of information comes out of the BH. And so,  using Eq.~(\ref{inforef}), we see that this happens when
 $\;N \simeq  \sqrt{S_{BH}(0)}\;$, which is the same as the coherence time.
On the other hand,   the transparency time $t_{trans}$ occurs when $\;dI/dS_{H}= 1\;$.  From  Eq.~(\ref{inforate}),
this transpires when $\;N C_{BH}\simeq1\;$, which is indeed the Page time.

Another clue is found by looking at the purity of the density matrix,
\be
P(\rho^{(N)})\equiv\frac{{\rm Tr} \left[\left(\rho^{(N)}\right)^2\right]}{\left({\rm Tr}\; \rho^{(N)}\right)^2} \;\simeq\;
\frac{1}{N}  {\rm Tr}\;\rho_H^2\left(1+ N C_{BH} \frac{ {\rm Tr}\left[\Delta\rho_{OD}\right]^2 }{ {\rm Tr}\;\rho_H^2} \right)\;.
\label{ratio}
\ee
The smallness of this ratio implies that the density matrix is still close to thermal, even at the Page time. However, the Page time appears
to be the moment when deviations from thermality are  starting to become significant, just as Page had asserted.

Inspecting  the purity, one can see that the radiation is already close to pure when $\;C_{BH}\lesssim 1\;$. Unlike the previous calculation of the information rate, which entailed expanding out a logarithm, Eq.~(\ref{ratio}) is reliable also for values of $\;N C_{BH} \gg 1\;$  provided that $\;C_{BH}< 1\;$. Hence, we can conclude that the radiation does parametrically  purify.

\section{Time-dependent radiation density matrix}

\subsection{A model of time-dependent emission}

We will now provide a more accurate account of the  back-reaction of the emitted particles.
Let us begin by assigning  a time-dependent wavefunction to the shell. Then both the mean position of the shell and its width could, in principle, become time dependent. Specifically, $R_S$ and  $C_{BH}$  are now  both functions of time,
\be
\left.\Psi_{shell}(R_{shell})\right|_{R_{shell}\to R_{S}(t)}\; = \;{\cal N}(t)^{-1/2} e^{-\frac{\left({R}_{shell}-R_{S}(t)\right)^2}{2C_{BH}(t) \left(R_{S}(t)\right)^2}}
\label{wavetime}
\;.
\ee
However, in this particular case, the width
$\;C_{BH}(t) \left(R_{S}(t)\right)^2=l_p^2/\pi\;$ is actually
a  constant.

What is required is the wavefunction in terms of $v$. For this, we use
\be
v_0-v_{shell}(t)\simeq \frac{1}{R_S(t)}\left( R_{shell}-R_S(t)\right)\;.
\label{vvRR}
\ee
Here, the time dependence  of $v_{shell}(t)$ is classical and due only to the classical time dependence of $R_S(t)$.

The resulting wavefunction is
\be
\left.\Psi_{shell}(v_{shell}(t))\right|_{v_{shell}(t)\to v_0}\;=  \;{\cal N}(t)^{-1/2} e^{-\frac{(v_0-v_{shell}(t))^2}{C_{BH}(t)}}\;.
\label{vVR}
\ee
For this parametrization, the width is time dependent.

It is more convenient to use the number of emitted particles $N$ as our time coordinate rather than $t$ or $v$. The Stefan--Boltzmann law for black-body emission leads to $\;N(t) = S_{BH}(0)\frac{t^{2/3}}{\tau^{2/3}}\;$, where $\;\tau=640S_{BH}(0)R_S(0)\;$ is the BH lifetime. We  will use  $N_{T}$ to denote the total number of particles emitted by a certain time, so that the multi-particle matrix is now an $N_{T} \times N_{T}$ block matrix. The time of emission of specific particles will be denoted by $N$, $N'$, {\em etc.}~. Of course, $\;N\leq N_T\;$.

\subsection{The time-dependent density matrix}

We can further  improve on the previous results by taking into consideration that the time of emission differs for the different  Hawking particles. In particular, the phase discontinuities associated with the logarithm in Eq.~(\ref{betaH}) depend on these emission times. This effect is not relevant  to the classical Hawking calculation but could be relevant to the phases of the semiclassical $\beta$ coefficients; {\em cf}, Eq.~(\ref{betaQ}). This is because the  shell-crossing time $v_{shell}(t)$ is different for different modes due to the shell continually depleting its mass; {\em cf}, Eq.~(\ref{vvRR}).

Now suppose that a given particle is emitted at ``time'' $\;N^{\prime}\;$ and another at $\;N^{\prime\prime}$.
Then the density matrix of Eq.~(\ref{rhoSCD}) should be replaced with
\bea
&&\rho_{SC}(\omega,\wt{\omega};N_T;N^{\prime},N^{\prime\prime})=
\int\limits_{-\infty}^{v_0}dv \int\limits_0^\infty d\omega'\int\limits_0^\infty  d\omega'' \frac{1}{2\pi} e^{iv(\omega^{\prime}-\omega^{\prime\prime})} \cr && \times \langle\Psi_{shell}(v_{shell}(N_T))| \beta^{\ast}_{\omega^{\prime}\omega,~SC}(N^\prime)
\beta_{\omega^{\prime\prime}\widetilde{\omega},~SC}(N^{\prime\prime})
|\Psi_{shell}(v_{shell}(N_T))\rangle\;.\;\;
\label{rhoSCT}
\eea
The density matrix depends on the three times $N^{\prime}$, $N^{\prime\prime}$ and $N_T$. The width of the wavefunction at $N_T$ controls the fluctuations in $v_{shell}$ and is a property of the BH, while $N^{\prime}$ and $N^{\prime\prime}$ are the emission times of the specific particles and are intrinsic to the quantum matter fields.

The $N^{\prime}$, $N^{\prime\prime}$ dependence enters only through the $\beta$'s,
\be
\beta_{\omega,\omega^{\prime},~SC}(N^{\prime}) \sim e^{i \omega' v_{shell}(N')} \;.
\ee
The wavefunction, on the other hand, depends on $N_T$,
and so the density matrix depends on  additional phases
that are missed when it is evaluated at a common time
as in \cite{slowleak}. In particular,
\bea
&&\rho_{SC}(\omega,\wt{\omega};N_T;N^{\prime},N^{\prime\prime})=
\int\limits_{-\infty}^{v_0}dv \int\limits_0^\infty d\omega'\int\limits_0^\infty  d\omega''  \frac{1}{2\pi} e^{iv(\omega^{\prime}-\omega^{\prime\prime})} \cr &\times& e^{i\omega^\prime\left( v_{shell}(N_T)-v_{shell}(N^\prime)\right)  }
e^{-i\omega^{\prime\prime}\left( v_{shell}(N_T)-v_{shell}(N^{\prime\prime})\right) }
\cr &\times& \langle\Psi_{shell}(v_{shell}(N_T))| \beta^{\ast}_{\omega^{\prime}\omega,~SC}(N_T)
\beta_{\omega^{\prime\prime}\widetilde{\omega},~SC}(N_T)
|\Psi_{shell}(v_{shell}(N_T))\rangle\;. \ \
\label{tdrho}
\eea

The expectation value in the last line of  Eq.~(\ref{tdrho})  is
the same as that of  time-independent situation, so that the difference
between the treatments is in  the additional phase factors
in the second line.
These  phases can be re-expressed as
\bea
e^{-i\omega^\prime\frac{C_{BH}(N^\prime)}{2} (N_T-N^\prime) }\
e^{-i\omega^{\prime\prime}\frac{C_{BH}(N^{\prime\prime})}{2} (N_T-N^{\prime\prime})}\;.
\label{phasef}
\eea

The details of the calculation leading to the phase factor~(\ref{phasef}) and the  rest of the evaluation of $\rho_{SC}$ are relegated to the Appendix. The final time-dependent result is rather simple:  An additional ``suppression'' factor multiplying the time-independent matrix  of Eq.~(\ref{rhoscf1}),
\bea
\Delta\rho_{SC}(\omega,\wt{\omega};N_T;N^{\prime},N^{\prime\prime})
\;=\; D(N_T;N^\prime,N^{\prime\prime})\Delta\rho_{SC}(\omega,\wt{\omega};C_{BH}(N_T)) \;.
\label{rhoscSupp}
\eea
The suppression factor is given by
\be
D(N_T;N^\prime,N^{\prime\prime})=\frac{1}{2}\left( e^{-\frac{1}{4} \frac{\left[C_{BH}(N^\prime) (N_T-N^\prime)\right]^2}{ C_{BH}(N_T)}}+e^{-\frac{1}{4}\frac{\left[C_{BH}(N^{\prime\prime}) (N_T-N^{\prime\prime})\right]^2}{ C_{BH}(N_T)}}\right)\;.
\label{suppfac}
\ee

The expression in Eq.~(\ref{rhoscSupp}) for the semiclassical correction to the Hawking density matrix is limited in its validity to the region of parameter space when  $C_{BH}(N^\prime) (N_T-N^\prime)$, $C_{BH}(N^{\prime\prime}) (N_T-N^{\prime\prime})$ are small. These factors are indeed small for most emitted particles. They become order unity only when $N_T$ becomes of order $S_{BH}(0)$ and, additionally,  the differences $N_T-N^\prime$, $N_T-N^{\prime\prime}$  become of order $S_{BH}(0)$.

\subsection{The coherence time}

The semiclassical correction to the density matrix in Eq.~(\ref{rhoscSupp})
now contains an extra  suppression factor.
The contribution from  a particle emitted at time $N$ is
\be
D(N_T;N)=e^{-\frac{1}{4} \frac{\left[C_{BH}(N) (N_T-N)\right]^2}{ C_{BH}(N_T)}},
\label{expsup}
\ee
and so a  new time scale appears,
\be
 N_{coh}(N_T;N) \;=\; \frac{\sqrt{C_{BH}(N_T)}}{C_{BH}(N)} \;=\; \frac{S_{BH}(N)}{ \sqrt{S_{BH}(N_T)}}\;.
\label{coh}
\ee

We have identified   $\;N_{coh}(N_T;N)$ as  the coherence scale for particle emissions.
For emissions that occurred near the time $N_T$  when the density matrix is being evaluated, $\;N_T-N\lesssim N_{coh}\;$, the particles posses some degree of entanglement. On the contrary, for earlier emissions,  $\;N_T-N\gg N_{coh}\;$,
the emitted particles are disentangled.

In Schwarzschild time, this new  scale is parametrically equal to the coherence time
\be
\;t_{coh}=\frac{R^2_S}{l_p}\;.
\ee
For instance, at the Page time, $\;t_{coh}(t_{Page};t_{Page})= 640\sqrt{ \frac{\pi}{2}} \frac{R_S^2(0)}{l_p}\left(1 +{\cal O}\left[C^{1/2}_{BH}(0)\right] \right)\;$.

The coherence scale is also the time that it takes the BH to emit order of $\sqrt{S_{BH}}$ particles. Consider that
\begin{eqnarray}
 N_{coh}(N_T;N) & = & \frac{ S_{BH}(0)-N} {\sqrt{S_{BH}(0)-N_T}} \cr
& =&   \sqrt{S_{BH}(0)}\left[1-\frac{N-\frac{1}{2}N_T}{S_{BH}(0)}+ \cdots\right]\;,
\end{eqnarray}
where the dots stand for terms that are higher order in
$\frac{N_T}{S_{BH}(0)}$. The point being that, as long as $N$ is close
to $N_T$, the  corrections are subleading and  $\; N_{coh}(N_T;N)\simeq\sqrt{S_{BH}(0)}\;$ follows.

This new timescale $N_{coh}$ (or $t_{coh}$) is the central result of our paper. We use it in an extensive way in the following analysis and the rest of our results depend crucially on its existence.  The appearance
of $t_{coh}$ in our formalism is quite natural  for  the following reason:

Let us consider  the time over which  the wavefunction $\Psi_{shell}$  changes significantly. An inspection of Eq.~(\ref{wave}) indicates that this happens when the Schwarzschild radius shrinks by an amount  $\;\Delta R_S\sim -\sqrt{C_{BH}}R_S\sim -l_p\;$. Then, since $\;\Delta R_S =\frac{\partial R_S}{\partial t}\Delta t\sim-\frac{l_p^2}{R_S^2}\Delta t\;$, it follows that  $\;\Delta t\sim\frac{R_S^2}{l_p}=t_{coh}\;$. Hence, the coherence time means the interval over which the overlap of the wavefunction at different times becomes small. The fact that $\;N_{coh}\ll S_{BH}\;$ ($t_{coh}\ll \tau_{BH}$)
can be attributed to  the width of the wavefunction being much smaller than the Schwarzschild radius or, equivalently, to the BH being semiclassical,
$\;C_{BH}\ll 1\;$.

\subsection{A simplified qualitative description of BH evaporation}

Let us start at time $\;N_T=0\;$ and follow the evaporation  for one interval  of the coherence time, $\;N^{(0)}\equiv N_{coh}(0;0)=\sqrt{S_{BH}(0)}\;$. This will define a block of size $\;\sqrt{S_{BH}(0)}\times \sqrt{S_{BH}(0)}\;$ in the multi-particle matrix. We then start the clock over at time $\;N_T= N^{(0)}\;$ and pretend that this is a newly born BH of smaller size (the original BH minus the first block). We  again follow the evaporation for a time  set by the coherence scale, but with the scale now  determined by this smaller-sized BH. That is,
$\;N^{(1)}\equiv N_{coh}(N^{(0)};N^{(0)})
=\sqrt{S_{BH}(N^{(0)})}\;$.
Then, by continually repeating this process,  we can parse the matrix  into
about $N_T/\sqrt{S_{BH}(0)}$ square blocks that are roughly of size $\sqrt{S_{BH}(0)}\times \sqrt{S_{BH}(0)}$ (although
each additional block is slightly smaller than the previous one).

The difference between Hawking's original description of BH evaporation and
ours is that, for  Hawking's picture, there are
$N_T$ blocks of size 1,  as  each emitted particle is independent of all  the other particles. Conversely, for  our previous
time-independent treatment, there is a single  block of size $N_T\times N_T$.

To a good approximation, each block can be viewed as the evolution of a newly born BH for its coherent  time scale, as the regions of the density matrix external to any block are highly suppressed. Then, as long as we proceed one block at a time, the suppression factor can be ignored.  All particles in the same block are approximately coherent and indistinguishable. Hence, the results of our time-independent treatment can be  applied.

Let us  now consider the effective expansion parameter
for a block that is ``born'' at a time $N_T$ which is  not too late in the evaporation process. By construction, the total particle number
of the block  is
about the same as the number of correlated particles at this time,  $N_{coh}(N_T;N)\simeq\sqrt{S_{BH}(N_T)}$
(here, $N$ means a time close to $N_T$). The  classicality parameter is approximately $C_{BH}(N_T)=S_{BH}^{-1}(N_T)\; $  because $C_{BH}$ evolves very slowly --- $\;\partial_{N_T}C_{BH}(N_T)=C_{BH}^2(N_T)\;$ --- except near the end of the evaporation. The effective expansion  parameter is then the product
 $\;N_{coh}(N_T;N)C_{BH}(N_T)\simeq \sqrt{C_{BH}(N_T)}\;$,
 which is obviously  less than one until the BH reaches the Planck scale.
 To compare,  the effective expansion parameter for the time-independent treatment   is $N_T C_{BH}(N_T)$, which is much larger than that of the block picture.

Let us next  determine the ``time of last block emission'' $N^{\ast}$. This
is the time when the number of particles remaining to be emitted $S_{BH}(0)-N^*\;$ is equal to the coherence time,
\be
 S_{BH}(0)-N^{\ast}\; =\; N_{coh}(N^{\ast};N)\;,
\label{lastblk}
\ee
where  $N$ is again  a  time close to $N^{\ast}$.  For future reference,  since $\;C_{BH}(N^{\ast})=(S_{BH}(0)-N^{\ast})^{-1}\;$,   Eq.~(\ref{lastblk}) is equivalent to the condition
\be
N_{coh}(N^{\ast};N)\ C_{BH}(N^{\ast})\;=\;1\;.
\label{ncbh}
\ee

Now, approximating  $N_{coh}(N^{\ast};N)$ by $\;N_{coh}(N^*;N^{\ast})=\sqrt{S_{BH}(N^{\ast})}=\sqrt{S_{BH}(0)-N^{\ast}}\;$, we find that the condition in Eq.~(\ref{lastblk}) becomes $\;S_{BH}(0)-N^{\ast}\;=\sqrt{S_{BH}(0)-N^{\ast}}\;$. This implies that the last block
consists of a single particle, which  does not make sense. As shown below, this is an indication that the block picture has broken down and the approximation $\;N_{coh}(N^{\ast};N)\simeq\sqrt{S_{BH}(0)-N^{\ast}}\;$
has become invalid by this  time.

To see this, let us use  Eq.~(\ref{coh}) to rewrite Eq.~(\ref{lastblk})  as
\be
 S_{BH}(0)-N^*\; =\; \frac{S_{BH}(N)}{\sqrt{S_{BH}(0)-N^*}}\;,
\label{lastblk1}
\ee
from which it follows that
\be
S_{BH}(0)-N^*\;=\;S^{2/3}_{BH}(N)\;=\;\left(S_{BH}(0)-N\right)^{2/3}\;.
\label{stransx}
\ee
We  now use another approximation which will turn out to be the correct way to estimate the emission time of the last block.
Expanding  the right-hand side of Eq.~(\ref{stransx}),  we have
\bea
S_{BH}(0)-N^*  &=& S^{2/3}_{BH}(0)\left(1 - \frac{2}{3} \frac{N}{S_{BH}(0)} + \cdots\;\right).
\eea
Then, since $\;\frac{N}{S_{BH}(0)} \lesssim 1\;$,
\be
S_{BH}(0)-N^*\;\simeq\;S^{2/3}_{BH}(0)\;,
\label{strans}
\ee
which is satisfied at the transparency $t_{trans}$  (as will be made explicit in Subsection~4.2). An equivalent form of Eq.~(\ref{strans}) is $\;C_{BH}(N^{\ast})\simeq S^{-2/3}_{BH}(0)\;$.

The interpretation of  the contradiction between the two approximations is clear; the block picture breaks down at time  $t_{trans}$, which happens to be one interval of the coherence time  before the end of evaporation (as also  made explicit in Subsection~4.2). At this point in time, the BH is still large and semiclassical, although its Schwarzschild radius is parametrically smaller than the initial radius $R_S(0)$ and the size of the remaining block is parametrically larger than $\sqrt{S_{BH}(0)}$.

Since the formalism of Section~2 can be applied to the block picture before it breaks down, we can estimate some associated quantities. For instance, the von Neumann entropy of a block that is  born at $N_T$ is  ({\em cf}, Eq.~(\ref{staylor}))
\be
S_{block}(N_T)\;\simeq\;
\sqrt{S_{BH}(N_T)}\left(1- \frac{1}{2} K\frac{1}{\sqrt{S_{BH}(N_T)}}\right)\;,
\label{sblock}
\ee
where we have also used that the thermal entropy of a block is approximately the same
as its particle number.

More interesting is the information output per  block. According to Eq.~(\ref{inforate}) and the above observations, the rate is
\be
\frac{dI_{block}}{dN}\;\simeq \;
\frac{K}{\sqrt{S_{BH}(N_T)}}\;;
\ee
meaning that, over the ``lifetime'' of the block,
\be
\Delta I_{block} \;\simeq \; K\;.
\label{iblock}
\ee
That is, each block emits about one bit of information.  This can also be seen directly from Eq.~(\ref{sblock}).

As there are roughly $\sqrt{S_{BH}(0)}$ blocks in total, the implication of the above simplified picture  is that only $\;\Delta I_{BH}\simeq \sqrt{S_{BH}(0)}\;$  ever gets released. However, this is incorrect because, even besides the break down at $t_{trans}$, the independent block picture is not perfectly accurate. The blocks overlap, correlations get built and accumulate. To pick up the information that comes out, one has to monitor the BH continuously, otherwise the information gets lost after each coherence time.

\section{Time dependence of information release}

This section  will focus on how the suppression factor and coherence scale impact upon  the purification of the density matrix and the transfer of
information.

\subsection{Time-dependence of the purity}

The purity of the  density matrix $\rho^{(N_{T})}\equiv \rho_{SC}(\omega,\wt{\omega};N_T;N',N'')$ is determined by the ratio
$\;P(\rho^{(N_T)})=\frac{{\rm Tr} \left[\left(\rho^{(N_T)}\right)^2\right]}{\left({\rm Tr}\; \rho^{(N_T)}\right)^2}\;$, which will be calculated next.
This result  will be the initial step towards distinguishing  the different phases of information release.

We first re-express  the multi-particle density matrix of Eq.~(\ref{nmatrix}) but with the  suppression factor now included.
In terms of the variables $N^{\prime}$, $N^{\prime\prime}$, each ranging from $0$ to $N_T$  and
with frequency labels suppressed, this is
\be
\rho_{SC}(N_T;N',N'')\;=\; \frac{1}{N_T}
\rho_H \delta_{N',N''}
\;+\; \frac
{C^{1/2}_{BH}(N_T)}{N_T}\Delta\rho_{OD}\;
D(N_T,N^\prime,N^{\prime\prime})
\left(1-\delta_{N',N''}\right)\;.
\label{nmatrix2}
\ee

As the matrix has already been  normalized  to yield unit trace, we need only calculate ${\rm Tr}\left[(\rho^{(N_T)})^2\right]$. Moreover, we do not have to consider the diagonal contributions because these will contribute at order $N_T^{-1}$  and do not ``mix'' with off-diagonal terms as far as this trace is concerned (see \cite{slowleak}). Hence, for current purposes, we can consider a simplified matrix for the off-diagonal correction,
\be
\rho_{OD}(N_T;N',N'')\;=\; \frac{C^{1/2}_{BH}(N_T)}{2N_T}\Delta\rho_{OD}
\;\left(e^{-\frac{1}{4} \frac{\left(N_T-N^\prime\right)^2}{ N^2_{coh}(N_T;N')}}
\;+\;e^{-\frac{1}{4} \frac{\left(N_T-N^{\prime\prime}\right)^2}{ N^2_{coh}(N_T;N^{\prime\prime})}}
\right)\;,
\label{simple}
\ee
where the exponents in Eq.~(\ref{suppfac}) for $D$
are now expressed  in terms of $N_{coh}$.

Since $N_T$ is  large, we can treat the discrete arguments of the density matrix as continuous. Now consider that
\be
\hspace*{-3in}{\rm Tr} \Big[\left(\rho_{OD}(N_T;N',N^{\prime\prime})\right)^2\Big] \nonumber \ee
\bea
&=& \int\limits^{N_T}_0dN^{\prime}\int\limits^{N_T}_0 dN^{\prime\prime} \int\limits^{N_T}_0 dN^{\prime\prime\prime} \;
 \rho_{OD}(N_T;N',N^{'''})
\rho_{OD}(N_T;N^{{\prime\prime\prime}},N^{\prime\prime})\delta(N'-N^{\prime\prime})\nonumber \\
&=& \int\limits^{N_T}_0 dN' \int\limits^{N_T}_{0} dN^{\prime\prime}\;
 \left[\rho_{OD}(N_T;N',N^{\prime\prime})\right]^2 = \frac{C_{BH}(N_T)}{4 N_T^{\ 2}} {\rm Tr} (\Delta\rho_{OD})^2\ {\cal I}\;,
\label{rhoOD2}
\eea
where ${\cal I}$ is given by
\be
{\cal I} \; = \;   \int\limits^{N_T}_0 dN' \int\limits^{N_T}_{0} dN^{\prime\prime}\;
\Big[e^{-\frac{1}{2} \frac{\left(N_T-N^\prime\right)^2}{ N^2_{coh}(N_T;N')}}
\;+\;e^{-\frac{1}{2} \frac{\left(N_T-N^{\prime\prime}\right)^2}{ N^2_{coh}(N_T;N^{\prime\prime})}}
\;+\; 2\;e^{-\frac{1}{4} \frac{\left(N_T-N^\prime\right)^2}{ N^2_{coh}(N_T;N')}}
\;e^{-\frac{1}{4} \frac{\left(N_T-N^{\prime\prime}\right)^2}{ N^2_{coh}(N_T;N^{\prime\prime})}}\Big]
\;.
\label{calI2}
\ee

Recalling that the suppression factors restrict $N'$, $N^{\prime\prime}$ to take on values close to  $N_T$ and that $C_{BH}(N)$ is a slowly varying function except at late times, we can make the approximation  $\;C_{BH}(N')$, $C_{BH}(N^{\prime\prime})= C_{BH}(N_T)\;$ and then evaluate the Gaussian integrals. For instance,
\bea
\int\limits^{N_T}_0 dN'\;e^{-\frac{1}{2} \frac{\left(N_T-N^\prime\right)^2}{ N^2_{coh}(N_T;N')}} &=&
\int\limits^{N_T}_0 dx\;e^{-\frac{1}{2}\frac{x^2}{N^2_{coh}(N_T;N_T)}}
\nonumber \\
&=& \sqrt{\frac{\pi}{2}}N_{coh}(N_T;N_T)\;,
\eea
where $N_T\gg 1$ has also been used  to treat the upper boundary
of the $x$ integral as infinite.

In this way, one ends up with
\be
{\cal I}\;=\;\sqrt{2\pi}N_TN_{coh}(N_T;N_T)+{\cal O}\left( C_{BH}^{-1}(N_T)\right)\;.
\ee
Then, reinserting the other factors from $(\rho^{(N_T)}_{OD})^2$
and dropping the subleading term, we arrive at
\be
{\rm Tr} \left[(\rho^{(N_T)}_{OD})^2\right]\;=\;\frac{\sqrt{2\pi}}{4}
\frac{N_{coh}(N_T;N_T)C_{BH}(N_T)}{N_T}
{\rm Tr}
\left[(\Delta \rho_{OD})^2\right]\;.
\label{trrho2}
\ee

As the purity of the Hawking matrix is $1/N_T$,  the purity of $\rho^{(N_T)}_{OD}$ is smaller by a factor of $\;N_{coh}C_{BH}\simeq C^{1/2}_{BH}\ll 1\;$. It appears that the purity of the off-diagonal correction only catches up to the small  purity of the Hawking matrix when $\;C_{BH}(N_T)\simeq 1\;$; implying that there is no chance for purification. However, it will be shown later on that such a conclusion is premature.

\subsection{The rate of information transfer}

It is interesting to compare the preceding calculation for ${\rm Tr}\left[(\rho^{(N_T)}_{OD})^2\right]$ with that of
our earlier study \cite{slowleak}.
We previously  obtained  $\;{\rm Tr} \left[(\rho^{(N_T)}_{OD})^2\right]\sim C_{BH}(N_T)\;$, so that the modified result in Eq.~(\ref{trrho2}) is smaller
by a factor of $\;\left[N_TC^{1/2}_{BH}(N_T)\right]^{-1}\simeq C^{1/2}_{BH}\;$. This estimate can be substantiated  as follows:  $N_T$ and $\;S_{BH}=C^{-1}_{BH}\;$  are
parametrically equal  for a ``typical BH", meaning  for times $\;t_{1bit}< t <t_{trans}\;$. In which case,  the time-dependent model effectively replaces $N_T$ with
$\;N_{coh}(N_T;N)= S_{BH}(N)\ C^{1/2}_{BH}(N_T)\simeq N_T C^{1/2}_{BH}(N_T)\;$.
Much in the same way, our previous time-independent estimates for the rate of information transfer can be corrected for time dependency
by replacing  $N_T$ with  $\;N_{coh}(N_T;N)\simeq N_TC_{BH}^{1/2}
\simeq  C_{BH}^{-1/2}(N_T)\;$
where appropriate. Here and below, $N$ means a time close enough to $N_T$ for
insignificant suppression.

For instance, consider the estimate for the information $I$ in Eq.~(\ref{inforef}). It should now be modified,
\bea
I(N_T)&=& S_H(N_T)-S(N_T) \cr  &\simeq& \frac{\wt K}{2} S_H(N_T) N_{coh}(N_T;N)\ C_{BH}(N_T)\;,
\label{inforatet}
\eea
with a numerical factor modifying $K$ to $\wt K$.

We can  use Eq.~(\ref{inforatet}) to determine when the first bit of information
comes out of the BH. For such early times,
\bea
&& \frac{\wt K}{2} S_H(N_T) N_{coh}(N_T;N)\ C_{BH}(N_T) \;\simeq\; \frac{\wt K}{2}S_H(N_T)\ C_{BH}^{1/2}(N_T)\;,
\label{inforatet2}
\eea
so the first bit of information comes out when  $\;S^{-1}_H(N_T) \simeq C_{BH}^{1/2}(N_T)\;$ or  $\;N_T\simeq \ S^{1/2}_{BH}(N_T)\;$.
This happens at the coherence time. Of course, we already knew this, since
each coherence-sized  block contains one bit of information; {\em cf}, Eq.~(\ref{iblock}). Hence,  $\;t_{1bit}= t_{coh}\;$, the same as for the previous
time-independent treatment.

It will be  shown below (also see
Eq.~(\ref{ncbh})) that the transparanecy  time
 occurs when  $\;N_{coh}(N_T;N)\ C_{BH}(N_T) \simeq 1\;$. This and Eq.~(\ref{inforatet}) tells us that the amount of information released by this time is
\be
I(t_{trans})\;\simeq\; \frac{\wt K}{2}S_H(N_T)\;.
\label{itrans2}
\ee
The value of $S_H$ by that time is parametrically equal to the total entropy of the BH, $S_H(t_{trans}) \simeq S_{BH}(0)$. So, parametrically, all the BH information is released by $t_{trans}$.

Another useful approximation for the information $I$ that is valid up to the transparency time is the following:
\bea
 I(N_T)&\simeq&\; \frac{\wt K}{2} S_H(N_T) N_{coh}(N_T;N)\ C_{BH}(N_T) \cr &\simeq& \frac{\wt K}{2} S_H(N_T)\ \frac{\sqrt{C_{BH}(N_T)}}{C_{BH}(N)}\ C_{BH}(N_T) \cr &\simeq& \frac{\wt K}{2} S_{H}(N_T)\  S_{BH}(0)\ C^{3/2}_{BH}(N_T)\;,
\label{inforate3}
\eea
with the last relation resulting from the approximation $\;C_{BH}(N)\sim C_{BH}(0) = S^{-1}_{BH}(0)\;$. Equation~(\ref{inforate3}) correctly estimates the value of the released information up to $t_{trans}$. Comparing with Eq.~(\ref{itrans2}), we see that the transparency time coincides with
$\; C_{BH}(N_T)\simeq S_{BH}^{-2/3}(0)\;$, as already claimed in Subsection~3.4
(see below Eq.~(\ref{strans})).
Notice, however, that the derivative $\frac{dI}{d S_H}$ cannot be estimated correctly from this expression because $S_{BH}(0)$ is a constant.

Let us next consider the modified version of Eq.~(\ref{inforate}),
which is obtained by differentiating  Eq.~(\ref{inforatet}). To evaluate the derivative $\left.\frac{dI}{dS_H}\right|_{N_T}$,  we recall that  $\;\partial_{S_H} N_T\simeq 1\;$,  $\;\partial_{N_T} C_{BH}\simeq C^2_{BH}\;$. The latter
can be ignored to leading order, and so
\be
\left.\frac{dI}{dS_H}\right|_{N_T}\;\simeq \;
\frac{\wt K}{2} N_{coh}(N_T;N)\ C_{BH}(N_T)\;,
\label{inforate2}
\ee
from which it is evident that the information transfer rate is
initially small but becomes order unity at the late stages of evaporation.

We have defined the transparency time as the moment at  which the rate of information transfer is unity
\be
\left.\frac{dI}{dS_H}\right|_{t_{trans}}\;\simeq \;1\;.
\ee
And so $t_{trans}$ is the time at which $N_{coh}(N_T;N)\ C_{BH}(N_T)\simeq 1\;$  as already stated. Let us recall that $t_{trans}$ has replaced  the Page time in this respect. We again see that there is nothing particularly special about the original Page time in our updated framework.

We now want to verify that the transparency time occurs at
one coherence time before the end of evaporation.
By this time, the BH still has an entropy
of $\;S_{BH}(t_{trans})= C^{-1}_{BH}(t_{trans})\simeq
S^{2/3}_{BH}(0)\;$.
And so we  start by setting  $\;[\Delta N]_{trans}  = S_{BH}^{2/3}(0)\;$,
where $\;[\Delta N]_{trans}=S_{BH}(0)-N_{trans}\;$ is  the time from transparency to evaporation.  In integral form, this is
\be
\int\limits^0_{[\Delta N]_{trans}} dN \;=\; -\frac{2\pi}{\hbar G} \int^0_{[\Delta R_S]_{trans}}  dR_S \; R_S  \;=\;S_{BH}^{2/3}(0) \;,
\ee
with the first equality following from $\;\frac{\partial N}{\partial R_S}=-\frac{2\pi R_S}{\hbar G}\;$.

Integrating, we then have $\;[\Delta R_S]_{trans}\simeq l_p S_{BH}^{1/3}(0)\simeq \left(\frac{l_p}{R_S(0)}\right)^{1/3} R_S(0)\;$.
But, since $\;\frac{dR_S}{dt}\simeq -\frac{l_p^2}{R^2_S}\;$, it follows that
$\;[\Delta t]_{trans}\simeq \frac{[\Delta R_S]^3_{trans}}{l_p^2}\simeq \frac{R_S^2(0)}{l_p}\;$.
That is, $\; [\Delta t]_{trans}\simeq t_{coh}\;$ as claimed.

The results of this section are summarized in Figure~1, showing the rate of information release.

\begin{figure}[t]
\ \hspace{1in}\scalebox{1.2} {\includegraphics{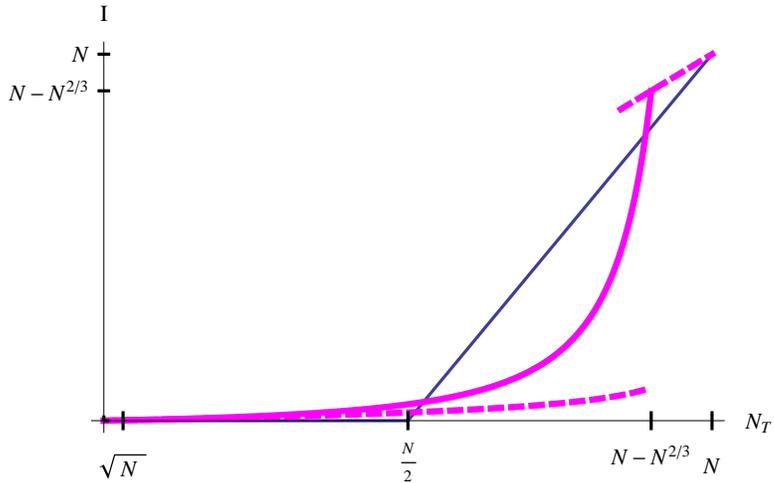}}
\caption{Information released as a function of the number $N_T$ of emitted Hawking particles shown for the Page model (blue) and our model (solid and dashed purple). Here, $N$ denotes the maximal value of $N_T$ which is approximately $S_{BH}(0)$. The lower dashed line depicts the  block picture of $\S$3.4. The upper dashed line is for $N_{coh} C_{BH}=1$, so it estimates the derivative at the transparency time correctly and corresponds to the approximation in Eq.~(\ref{inforate2}). The solid purple line depicts the approximation in Eq.~(\ref{inforate3}).}
\end{figure}

\subsection{Qualitative discussion of the final purification}

Let us now address the question of  what happens at times later than $t_{trans}$, when the BH becomes parametrically Planckian in size,
$\;S_{BH}(N_T)\gtrsim 1\;$.
Our results are not  formally  valid in this region of parameter space, as
indicated by the derivation of the suppression factor in the Appendix. Although a more rigorous analysis will eventually be required, we argue  that the scaling of $\;N_{coh}(N_T;N)$  does capture its correct behavior even in the region where our analysis cannot be formally validated.
Based on the scaling of $\;N_{coh}(N_T;N)$, which indicates that $\;N_{coh}(N_T;N)\to S_{BH}(N)\;$ in the late-time limit, we will argue that the radiation does indeed purify at the late stages of the evaporation.

Let us substantiate our arguments by looking at the relevant
integral, which is that of Eq.~(\ref{calI2}) with
$\;C_{BH}(N_T)\lesssim 1\;$ meant as a number of order unity but still smaller than 1,
\bea
{\cal I} \; &=& \;   \int\limits^{N_T}_0 dN' \int\limits^{N_T}_{0} dN^{\prime\prime}\;
\Big[e^{-\frac{1}{4} \frac{2}{C_{BH}(N_T)} \left[C_{BH}(N^\prime) (N_T-N^\prime)\right]^2}+e^{-\frac{1}{4}\frac{2}{C_{BH}(N_T)}\left[C_{BH}(N^{\prime\prime})(N_T-N^\prime)\right]^2}
\nonumber \\ &&
+ 2\;e^{-\frac{1}{4} \frac{1}{ C_{BH}(N_T)} \left[C_{BH}(N^\prime) (N_T-N^\prime)\right]^2}\;e^{-\frac{1}{4}\frac{1}{ C_{BH}(N_T)} \left[C_{BH}(N^{\prime\prime}) (N_T-N^{\prime\prime})\right]^2}\Big]
\;.
\eea
The magnitude of any of the exponents is at most
of order unity. For instance, setting $\;N'=1\;$ ({\em i.e.}, the initial
emission of radiation),  one finds that the first exponent goes as
$\;\frac{1}{4} \frac{2}{C_{BH}(N_T)} \left[C_{BH}(1)(N_T-1)\right]^2 \simeq
\frac{1}{4} \frac{2}{C_{BH}(N_T)}\left[C_{BH}(0)S_{BH}(0)\right]^2=\frac{1}{4} \frac{2}{C_{BH}(N_T)}\simeq 1\;$. Clearly, for $N^{\prime}$, $N^{\prime\prime}>1$, similar estimates are also valid. Hence,  the Gaussians are turning into Heaviside functions and, parametrically, $\;{\cal I}\sim 4 (N_T)^{2}\;$.

Then, using the estimate $\;{\cal I}\sim 4 (N_T)^{2}\;$ in Eq.~(\ref{rhoOD2}), we have
\be
{\rm Tr} \left[(\rho_{OD}^{(N_T)})^2\right]\;\sim\;
{C_{BH}(N_T)}
{\rm Tr}
\left[(\Delta \rho_{OD})^2\right]\;\sim\; 1\;,
\ee
from which $\;P(\rho^{(N_T)})\sim 1\;$ follows.
The interpretation is that the density matrix does
appear to have purified towards the end of the evaporation. We expect to provide
a more rigorous analysis of the late-time purification at a later  time \cite{future}.

The purification of the density matrix can be attributed to the late-time scaling $\;N_{coh}(N_T;N)\simeq S_{BH}(N)\;$, which implies that even the earliest emitted particles are part of the coherent radiation. At a  first glance, this seems strange inasmuch as the dimensionless time scale $\;\Delta N=N_T-N\;$ and the dimensionless particle frequencies  are conjugally related ({\em cf}, Eq.~(\ref{tdnotnine})) in  such a way that both have widths going as $C_{BH}^{-1/2}$ (see  the relevant discussion in the Introduction). But this observation overlooks the fact that the coherence time depends on three different  time scales; the emission times of a given  pair of  particles and the evolution time of the collapsing  shell.  At late enough times  when the width of the wavefunction for the  shell grows to  order unity,  this distinction between time scales  becomes important. Nevertheless, addressing this question in a quantitative way will provide an interesting subject matter for our future work.

Another surprise is the apparent suddenness of the purification. After all, the ``action''  only seems to begin at $t_{trans}$, which is but one coherence time before the end. This is, to some extent, an artifact of the choice of time parameter; the evolution of the BH is more gradual when described in terms of the monotonically increasing classicality parameter $C_{BH}(N_T)$. As this parameter measures the degree of classicality of the geometry, one could  argue that it is also the most natural choice of clock for the current framework.

\section{Early-Late Entanglement}

Let us now address the entanglement between early and late-time  radiation, both for a ``typical'' BH and  for an  ``old'' one.  The results should be relevant for an eventual resolution of the firewall paradox \cite{AMPS}, as this
puzzle  is often posed as a conflict as to which subsystem the late radiation is entangled with and how strongly. Here, we will calculate the time dependence of the early--late entanglement  but  defer addressing the implications to the firewall paradox until a later article \cite{inprog}.

We will now  use the multi-particle  density matrix in explicit Dirac notation,
\bea
\rho_{SC}(N_T;N',N'') &=& \frac{1}{N_T}
\rho_H \delta_{N',N''}|N'\rangle\langle N''| \\
&+&
\frac{C^{1/2}_{BH}(N_T)}{2N_T}\Delta\rho_{OD}\;D(N_T;N',N'')
\left[1-\delta_{N',N''}\right]|N'\rangle\langle N''|\;,\nonumber
\label{nmatrix3}
\eea
where the suppression factor $D(N_T;N',N'')$ is defined in Eq.~(\ref{suppfac}).

\subsection{Entanglement for $t<t_{trans}$}

Let us first discuss the case that the BH is typical, namely, for times $t_{1bit}<t<t_{trans}$. First, we have to choose a reference time $N_{cut}$ and  factor the Hilbert-space $|N\rangle$  into the states of ``early emissions" $|N_E\rangle$, for which  $\;N_E\leq  N_{cut}\;$, and
``late emissions''  $|N_L\rangle$, for which $\;N_L\geq N_{cut}\;$.
In our framework, the natural choice of  ``cutoff'' is one coherence time
prior to $N_T$,
\be
N_{cut}=N_T-N_{coh}(N_{cut};N_{cut})=N_T-\sqrt{S_{BH}(N_{cut})}\;.
\label{ncut1}
\ee
At the end of evaporation, $N_{cut}$ is the transparency time.

The density matrix is then expressed on the product space $|N_E\rangle\otimes|N_L\rangle$,  $\;\rho_{E\otimes L} =\rho_{E\otimes L}(N_T;N_E',N_L',N_E'',N_L'')\;$. It   is given by
\bea
\label{earlylatem1}
&& \rho_{E\otimes L}(N_T;N_E',N_L',N_E'',N_L'')= \frac{1}{N_{prod}} \rho_H\otimes \rho_H |N_E'\rangle |N_L'\rangle\langle N_E''|
\langle N_L''|
\delta_{N_E',N_E''} \delta_{N_L',N_L''}\nonumber \\
&+& \frac{C_{BH}(N_T)}{4N_{prod}}
\Delta\rho_{OD}\otimes\Delta\rho_{OD} \cr &\times& \Big\{
 D(N_T;N_E',N_E'')
 D(N_T;N_L',N_L'')
|N_E'\rangle |N_L'\rangle\langle N_E''|
\langle N_L''|
_{(N_E'\neq N_E''\;,\;N_L'\neq N_L'')}\nonumber \\
&+&
 D(N_T;N_E',N_L'')
 D(N_T;N_E'',N_L')
|N_E'\rangle |N_L'\rangle\langle N_E''|
\langle N_L''|
_{(N_E'\neq N_L''\;,\;N_E''\neq N_L')}
\Big\}
\;,
\eea
where $\;N_{prod}=N_{cut}(N_T-N_{cut})\;$. The products
$\rho_H\otimes \rho_H$, $\Delta\rho_{OD}\otimes\Delta\rho_{OD} $
should be regarded as shorthand notation for
$\;\rho_{H}(\omega_{E^{\prime}},\widetilde{\omega}_{E''})\otimes\rho_{H}(\omega_{L^{\prime}},
\widetilde{\omega}_{L''})\;$,
$\;\Delta\rho_{OD}(\omega_{E^{\prime}},\widetilde{\omega}_{E''})
\otimes\Delta\rho_{OD}(\omega_{L^{\prime}},\widetilde{\omega}_{L''})\;$
for the first term
inside the curly brackets and
$\;\Delta \rho_{OD}(\omega_{E^{\prime}},\widetilde{\omega}_{L''})\otimes\Delta \rho_{OD}(\omega_{E''},\widetilde{\omega}_{L^{\prime}})\;$
for the second.

It is  the second term within the curly brackets that stores the entanglement between early and late radiation,
\be
\Big[ e^{-\frac{1}{4} \frac{\left(N_T-N_E^{\prime}\right)^2}{N^2_{coh}(N_T;N_E^{\prime})}}
\;+\; e^{-\frac{1}{4} \frac{\left(N_T-N_L^{\prime\prime}\right)^2}{N^2_{coh}(N_T;N_L^{\prime\prime})}}
\Big]
\;\times\;
\Big[ e^{-\frac{1}{4} \frac{\left(N_T-N_E^{\prime\prime}\right)^2}{N^2_{coh}(N_T;N_E^{\prime\prime})}}
\;+\; e^{-\frac{1}{4} \frac{\left(N_T-N_L^{\prime}\right)^2}{N^2_{coh}(N_T;N_L^{\prime})}}\Big]
|N_E'\rangle |N_L'\rangle\langle N_E''|
\langle N_L''|\;.
\label{keyterm}
\end{equation}
One can already see the source of entanglement; the density matrix does not factor into $\rho_E\otimes\rho_L$, rather there are correlations.

Let us now trace over the late radiation to obtain the reduced density matrix for the early radiation $\rho_E$. The trace over the late radiation of the Hawking term is calculated in a straightforward way and that of the first term within the curly brackets of Eq.~(\ref{earlylatem1}) trivially vanishes.
The only relevant term in Eq.~(\ref{earlylatem1}) is therefore the second term in the curly brackets. To evaluate it, we need to perform the following integral:
\be
{\cal J}_b\;=\; \int\limits^{N_T}_{N_{cut}} dN_L\;
e^{-\frac{1}{4}b \frac{\left(N_T-N_L\right)^2}{N^2_{coh}(N_T;N_L)}}
\;,
\label{calJ}
\ee
where $b$ is either 0, 1 or 2 depending on which of the four different
products of exponents in expression (\ref{keyterm})  is being considered.

It can be seen that, for a BH of typical age, the width of the Gaussian
in Eq.~(\ref{calJ}) $N_{coh}(N_T;N_L)$
is approximately the same as $N_{coh}(N_{cut};N_{cut})$
for admissible values of $N_L$. But
 $\;N_{coh}(N_{cut};N_{cut})=N_T-N_{cut}\;$ (see Eq.~(\ref{ncut1})),
meaning that the width spans over the range of integration.  Hence, $\;{\cal J}_b\simeq N_T-N_{cut}\;$ for  $\;b=0,1,2\;$.

Applying this estimate of ${\cal J}_b$, we then obtain a reduced density
matrix of the form
\bea
\rho_{E}(N_T;N_E',N_{E}'')&=& \frac{\rho_H}{N_{cut}}
| N_E'\rangle\langle N_E''|\delta_{N_E',N_E''} \nonumber \\
&+&
\frac{C_{BH}(N_T)}{4N_{cut}}\Delta\rho_{OD}^2(\omega_{E'},\omega_{E''})
\Big(1\;+\; e^{-\frac{1}{4} \frac{\left(N_T-N_E^{\prime}\right)^2}{N^2_{coh}(N_T;N_E^{\prime})}}
\Big)
\nonumber \\
&\times&\Big(1\;+\; e^{-\frac{1}{4} \frac{\left(N_T-N_E^{\prime\prime}\right)^2}{N^2_{coh}(N_T;N_E^{\prime\prime})}}
\Big)
| N_E'\rangle\langle N_E''|\;.
\label{redmattress}
\eea
Here, $\;\Delta\rho_{OD}^2(\omega_{E'},\omega_{E''})=\int d\omega \Delta\rho_{OD}(\omega_{E'},\omega)\Delta\rho_{OD}(\omega,\omega_{E''})\;$. Unlike $\Delta\rho_{OD}$, which is purely off-diagonal, $\Delta\rho_{OD}^2$ does have a diagonal component.

Let us now  consider times
$t_{1bit}<t<t_{trans}$.
In this case, the Gaussian-suppressed
terms are subdominant, leaving
\be
\rho_{E}(N_T;N_E',N_{E}'')\;=\; \frac{\rho_H}{N_{cut}}
| N_E'\rangle\langle N_E''|\delta_{N_E',N_E''}
+ \frac{C_{BH}(N_T)}{4N_{cut}}\Delta\rho_{OD}^2(\omega_{E'},\omega_{E''})
| N_E'\rangle\langle N_E''|\;.
\label{redmattress2}
\ee

The Gaussian suppression has disappeared and has been replaced by a
factor of $C_{BH}$ on the correction term.
The  von Neumann entropy per particle is given by (see
Footnote~1),
\be
\;\frac{S_{ent}}{N_{cut}}=-{\rm Tr}_E\left[\rho_E\ln{\rho_E}\right].\;
\label{vonNE}
\ee
In fact, to leading order, we need only consider
contributions from the diagonal elements of $\rho_{E}(N_T;N_E',N_{E}'')$.
This is because the large number of off-diagonal elements, a factor of $\sim N_{cut}$ more of these than diagonal ones, enters only at quadratic order, leading
to the additional suppression
$\;N_{cut}C^2_{BH}\ll C_{BH}\;$.

Now, if one uses the standard definition of entanglement for pure states and applies it to the  Hawking density matrix, it comes out as entangled. We know that the Hawking part of the matrix is thermal because of the tracing over the negative energy in-modes, and so it does not represent any entanglement between late and early radiation. Formally, one has to use an appropriate definition for mixed-state entanglement such as the  ``positive partial transpose'' criterion \cite{Peres,Horod}. Rather than use this sophisticated criteria, we will calculate the entanglement entropy and subtract from it the contribution from the Hawking density matrix.

We proceed  by expanding the logarithm of the density matrix of Eq.~(\ref{redmattress2})  in the von Neumann formula in Eq.~(\ref{vonNE}) to linear  order in $C_{BH}(N_T)$. Only the diagonal elements of $\rho$ contribute to this order, as just explained. We then subtract from the answer the zeroth order result coming
from the Hawking matrix. We also drop a factor of $\ln{N_T/2}$ that is due to the resolution of Gibbs' paradox for indistinguishable particles. The final result of this procedure is then
\be
S_{ent} \;=\; \frac{1}{2}N_{cut}C_{BH}(N_T){\rm Tr}\left[(-\ln{\rho_H})(\Delta\rho_{OD})^2\right]\;.
\label{entEL}
\ee
For a typical BH, $\;N_{cut}\simeq N_T\simeq C_{BH}^{-1}(N_T)\;$.
We can conclude that the entanglement entropy is of order unity, $\;S_{ent}\sim 1\;$.

Let us next consider the  entanglement  entropy at the transparency time, for which ({\em cf}, Subsection~4.2) $\;C_{BH}\simeq N_T^{-2/3}\;$ and
$\;N_{cut}\simeq N_T-S_{BH}^{2/3}(0)\simeq N_T\;$, so that
$\;S_{ent}\simeq N_T^{1/3}$. This is
significant compared to earlier times but well short
of that expected at the purification scale, $\;S_{ent}\simeq N_T\;$.
Hence, the time scale for  maximal entanglement must still  be  later than the transparency time.

\subsection{Qualitative discussion of the entanglement entropy for $t>t_{trans}$}

Our expectation is  that purification is indeed attained at the last phase of evaporation $t>t_{trans}$. Previously, we presented a qualitative argument based on the purity of the radiation density matrix. Here, we will discuss in a qualitative way the entanglement entropy at late times when $\;C_{BH}(N_T)$
approaches unity.  We hope to be able to present a more precise analysis in the future \cite{future}.

Let us begin by revising the form of the reduced density matrix.
We once again set $\;C_{BH}(N_T)\lesssim 1\;$ and,  because
the Gaussian suppression factors  become like theta functions
at late enough times ({\em cf}, Subsection~4.3),
replace the exponentials in Eq.~(\ref{redmattress}) with $1$'s.
Then
\be
\rho_{E}(N_T;N_E',N_{E}'')\;=\; \frac{\rho_H}{N_{cut}}
| N_E'\rangle\langle N_E''|\delta_{N_E',N_E''}
+  \frac{C_{BH}(N_T)}{ N_{cut}}\Delta\rho_{OD}^2(\omega_{E'},\omega_E'')
| N_E'\rangle\langle N_E''|\;.
\label{redmattress3}
\ee
In this case,  the ``correction'' term in $\rho_{E}$ is  the dominant one.
The Hawking contribution is ``only'' diagonal whereas the  correction
uniformly fills up the entire matrix.

Let us now recall that a uniform  $M\times M$
matrix filled with (say) $c$'s can  be diagonalized  to yield  a single non-vanishing
eigenvalue, $\;\lambda= c M\;$.
In this way,  the correction part of the matrix  can  be reduced
  to a diagonal matrix
with a single non-zero entry,
$\;\lambda = C_{BH}(N_T) {\rm Tr} \left[\Delta\rho_{OD}^2\right]\;$.
Once the Hawking contribution is  discarded, the entanglement entropy
can be calculated  in terms of the eigenvalue $\lambda$ given above,
\be
S_{ent}\; =\;  -\lambda N_{cut}\ln{\lambda}\;.
\ee
That is, a late-time entanglement of order $\;N_{cut}\simeq N_T\simeq
S_{BH}(0)\;$ as expected.

An order $N_T$ entanglement indicates a pure state while a small entanglement is an indication of a product state. Hence, the radiation does (parametrically) purify, but only in the final stages of the BH evaporation.

It is worth re-emphasizing that this conclusion should only be viewed as a
qualitative
one. We have greatly simplified matters by treating the Gaussian suppression
factors as theta functions in the late-time limit. In reality,
these late-time Gaussian factors are not exactly uniform. Nevertheless, the matrix is close enough to uniform to suggest
that our qualitative results  will survive a more  accurate treatment.

Let us further emphasize that the changing  coherence scale is the physical mechanism
which enables the entanglement between late and early modes to
(parametrically) maximize.
This entangled region  fills up a square block  of size
$N_{coh}(N_T)$
that is
typically of order $\sqrt{N_T}$ but grows to $N_T$ near the end of evaporation.
Meaning that,
at the end, the entangled region extends over the entirety of the Hilbert space for the emitted
particles.

\section{Conclusion}

Let us summarize our findings: We started by reviewing our previous calculation that improves Hawking's calculation of the density matrix for BH radiation by incorporating  the background quantum fluctuations. The novel feature of this semiclassical treatment is the presence of off-diagonal elements in the density matrix. We have then further improved our previous calculation of the density matrix by taking time dependence into consideration. The radiation is emitted  continuously and the geometry is continually evolving due to the back-reaction of these emitted particles.

Our main result is the discovery of the coherence time $t_{coh}$.  This scale affects the density matrix by introducing  an extra suppression factor in the off-diagonal elements that limits the extent in time over which emitted  particles are coherent. For most of the   BH evaporation process, this number of coherent particles is about $\sqrt{S_{BH}(0)}$, which is much
smaller than the total number of emitted particles during the lifetime of the
BH, $\;\sqrt{S_{BH}(0)}\ll S_{BH}(0)\;$. We have also identified a clear physical reason for the appearance of the coherence time:
The wavefunction of the BH at one time is  nearly orthogonal
to the wavefunction at another  when the time separation is $t_{coh}$, causing
the  emissions of particles that are separated by more than $t_{coh}$ to  become incoherent.

That some of the particle emissions are coherent is what allows for a unitary process of evaporation. In this way, the wavefunction is serving as the conduit for total information flow from the burning matter system to the final state of external  radiation. This conclusion was substantiated by three calculations: First, the trace of the square of the radiation density matrix  becomes larger at late times and parametrically approaches unity,  $\;{\rm Tr}(\rho^2)\sim 1\;$. Second,  the total information released by the BH is of the same order as its total information content, $\;I\sim S_{BH}(0)\;$. Third, the  late-time entanglement entropy between the early and late-emitted radiation is also of this order,  $\;S_{ent}\sim S_{BH}(0)\;$.

Qualitative arguments show that the number of coherent particles begins to grow rapidly one coherence time before the end of evaporation and spans the entirety of the emitted particles by the very end. This growth is surprising and deserves a more precise treatment. Evidently, the key to this mechanism is the wavefunction of the collapsing shell and the existence of several different time scales. This wavefunction provides a Gaussian width for the time lapse between particle emissions that depends on these particular emission  times  as well as the time scale in the evolution of the BH. The former scales are fixed by the geometry at the times of  emission, whereas the latter is changing as the  back-reaction  from the particles effectively  shrinks the shell.

We have identified the time-of-first-bit $t_{1bit}$ as occurring at a time  $t_{coh}$ after the emission of the first Hawking particle. On the other hand,  the transparency time $t_{trans}$, which is  the moment when  the rate of information flow reaches order unity,  occurs at a time $t_{coh}$ before the end of evaporation.  Finally, the purification
time  only occurs after $t_{trans}$ when the BH is still large but parametrically approaching Planckian dimensions. The Page time \cite{page}, which is attributed to the time of transparency in the Page model, no longer has any specific meaning in our framework.   It has been split into two different time scales $t_{1bit}$ and $t_{trans}$.  We expect that this distinction could be essential to resolving the recently posed firewall paradox \cite{AMPS}. For instance, let us suppose that a firewall is symptomatic of a transparent  BH, as is often implied to be the case.  Then our revised picture would delay the need for a firewall from the Page time to a parametrically smaller interval before the end of evaporation. This is a matter that we hope to report on in the near future \cite{inprog}.

The emerging picture of the phases of information release during BH evaporation is then the following: The first bit of information comes out from the BH after one coherence time. Then the information continues to come out of the BH at a nearly constant rate of 1 bit per coherence time until the transparency time is approached. By this time, the rate of information release becomes unity. The amount of information released by the transparency time is already of the order of the total entropy of the BH. After $t_{trans}$, our description of the BH radiation is only qualitative. But, based on the scaling of the quantities that could be calculated, we have argued that the radiation purifies quickly when the BH evaporation nears its final stages.

\section*{Acknowledgments}

We thank Sunny Itzhaki for discussions. The research of RB was supported by the Israel Science Foundation grant no. 239/10. The research of AJMM received support from an NRF Incentive Funding Grant 85353 and an NRF KIC Grant 83407. AJMM thanks Ben Gurion University for their hospitality during his visit.

\appendix

\section{Determining the Suppression Factor}

Our starting point here is Eq.~(\ref{tdrho}) and we focus on the impact of the additional phase factors that appear there,
\be
e^{i\omega^\prime\left( v_{shell}(N_T)-v_{shell}(N^\prime)\right)  }
e^{-i\omega^{\prime\prime}\left( v_{shell}(N_T)-v_{shell}(N^{\prime\prime})\right) }\;.
\label{pf1}
\ee

We will sometimes use a different choice of  variables and change from $N^\prime$, $N^{\prime\prime}$ to $\;\on=\frac{N^\prime+N^{\prime\prime}}{2} \;$,
$\;\delta N=\frac{N^{\prime\prime}-N^{\prime}}{2}\;$, so that
\bea
N^{\prime\prime}\;=\;\on+\delta N\;, \cr
N^{\prime}\;=\;\on-\delta N\;.
\label{npnpp}
\eea

We wish to express the phase factors in Eq.~(\ref{pf1}) in terms of
the particle number. For this, we define
\bea
\;\Delta N^{\prime}&=&N_T-N^{\prime}\; \cr
\;\Delta N^{\prime\prime}&=&N_T-N^{\prime\prime}\;.
\eea
We will  assume that the  differences $\Delta N^{\prime}$ and
$\Delta N^{\prime\prime}$ are small (in a sense made explicit below) and expand the phase factors accordingly. Our premise being that the expectation value on the left-hand side  of Eq.~(\ref{tdrho}) rapidly approaches zero for large enough $\Delta N^{\prime}$ and  $\Delta N^{\prime\prime}$. This assumption will be justified by its self-consistency.

We will expand the phases by  using a suitably  modified version of
Eq.~(\ref{vvRR}),
$\;v_0-v_{shell}(t)\simeq \frac{1}{R_S(0)}\left( R_{shell}-R_S(t)\right)\;$.
The point here is that the product of the dimensionless frequency and dimensionless advanced time does not depend on the time-dependent scale $R_S(t)$  that is used to make both dimensionless. Then, as our purpose is to  expand out the entire  phases and not just the $v$'s, the canceled-out factors of  $R_S(t)$ should not be included in the expansions.

The partial derivative $\frac{\partial R_S}{\partial N}$ is also required
and can be evaluated using the fact that
$\;N(t)=S_{BH}(0)-S_{BH}(t)={\rm const.}-\frac{\pi (R_S(t))^2}{\hbar G}\;$,
which gives us
\be
\frac{\partial R_S}{\partial N}\;=\; -\frac{\hbar G}{2\pi R_S}\;.
\label{expanv}
\ee
Hence,
\bea
v_{shell}(N_T)\;=\;v_{shell}(N^{\prime\prime})- \frac{C_{BH}(N'')}{2}\Delta N^{\prime\prime} +\cdots\;, \cr
v_{shell}(N_T)\;=\;v_{shell}(N^{\prime})- \frac{C_{BH}(N')}{2}\Delta N^{\prime}+\cdots \;,
\eea
where the $\cdots$ denote higher orders in $C_{BH}$. The second  expansion is  well defined provided that $\;C_{BH}(N^{\prime})\Delta N^{\prime} \lesssim 1\;$
(and similarly for the other one), which is equivalent to
\be
\Delta N^{\prime} \;\lesssim\;\frac{R^2_S(N')}{\hbar G}\;\sim \; S_{BH}(N')\;.
\ee
This is on the  order of the total number of Hawking particles that will be emitted during the whole period from $N^{\prime}$ to the end of the lifetime of the BH,
and so this restriction is a  weak one. We can  conclude  that the first-order
term in the expansions is a good approximation to the exact value
until $N_T$ becomes of order of $S_{BH}(0)$ and, then, is valid
if the  differences $\Delta N'$, $\Delta N''$ are smaller than $S_{BH}(0)$.

Evaluating the expectation value of Eq.~(\ref{tdrho}) in this way,
we obtain a modified form for the quantity $\wh{\Delta I}_{SC}(C_{BH}(N_T))$
that appears  in Eq.~(\ref{j6}),
\bea
&& \wh{\Delta I}_{SC}(N_T;N^{\prime},N^{\prime\prime}) \; = \; \wh{\Delta I}_{SC}(C_{BH}(N_T)) \cr &&  \times\; e^{-i\omega^\prime\left(\frac{C_{BH}(N^\prime)}{2} (N_T-N^\prime)\right)  }\
e^{i\omega^{\prime\prime}\left(\frac{C_{BH}(N^{\prime\prime})}{2} (N_T-N^{\prime\prime})\right)}
\;.
\label{tdnotnine}
\eea

To leading order in $C_{BH}$, it is sufficient to consider only the
explicitly shown exponential  phases. This is because the corrections to  other appearances of $R_S$ in  Eq.~(\ref{rhoSCT})  will pick up an overall factor
of $\;\frac{1}{R_S}\frac{\partial R_S}{\partial N} \simeq C_{BH}\;$
and the off-diagonal elements are already $\sim C_{BH}^{1/2}\;$.

We now understand how to modify $\Delta\rho_{SC}$ in Eq.~(\ref{rhosc})
to obtain its time-dependent  form,
\bea
&&\Delta\rho_{SC}(\omega,\wt{\omega};N_T;N^{\prime},N^{\prime\prime})
\;=\;
\frac{t^*_\omega t_{\wt{\omega}}}{(2\pi)^3}  \frac{C_{BH}(N_T)}
{(\omega \wt{\omega})^{1/2}} \Gamma\left(1+i2\omega\right)
\Gamma(1-i2\wt{\omega})
\ \hbox{\Large\em e}^{-\pi(\omega+\wt{\omega})} \cr &&
\times \;\int\limits_0^\infty d\omega'\int\limits_0^\infty  d\omega''   \ e^{-\frac{(\omega'-\omega'')^2}{4} C_{BH}(N_T)}
(\omega')^{-1/2-2 i{\omega}}
(\omega'')^{-1/2+2 i{\wt{\omega}}}\;
\cr &&  \times\; e^{-i(\omega^\prime-\omega^{\prime\prime})\frac{C_{BH}(\on)}{2} (N_T-\on) }\
e^{-i(\omega^\prime+\omega^{\prime\prime})\frac{C_{BH}(\on)}{2} \delta N }
\;,
\label{tdrho1}
\eea
where   $\;C_{BH}(N^{\prime})\simeq C_{BH}(N^{\prime\prime})\simeq C_{BH}(\on)\;$
has been employed.

We then need to evaluate the integral,
\bea
{\cal I}(\on,\delta N)&=&\int\limits_0^\infty d\omega'\int\limits_0^\infty  d\omega''   \ e^{-\frac{(\omega'-\omega'')^2}{4} C_{BH}(N_T)}
(\omega')^{-1/2-2 i{\omega}}
(\omega'')^{-1/2+2 i{\wt{\omega}}}\;
\cr &&  \times\; e^{-i(\omega^\prime-\omega^{\prime\prime})\frac{C_{BH}(\on)}{2} (N_T-\on) }\
e^{-i(\omega^\prime+\omega^{\prime\prime})\frac{C_{BH}(\on)}{2} \delta N }
\;.
\label{tdrho11}
\eea

Following  \cite{slowleak}, we change variables to $\;Y=\omega'-\omega''\;$ and $\;Z=(\omega'+\omega'')/Y\;$
\bea
{\cal I}(\on,\delta N)&=&
\int\limits_{0}^\infty d Y \ e^{-\frac{Y^2}{4}C_{BH}(N_T)}\
Y^{-i2(\omega-\wt{\omega})}\ e^{- i  Y  \frac{C_{BH}(\on)}{2}(N_T-\on)} \nonumber \\
&\times&\Biggl[
\int\limits_{1}^\infty  dZ \left(Z+1\right)^{-1/2-i2\omega}
\left(Z-1\right)^{-1/2+i2\wt{\omega}} e^{-i  Y Z \frac{C_{BH}(\on)}{2} \delta N}
\nonumber \\  &+& \int\limits_{1}^\infty  dZ \left(Z-1\right)^{-1/2-i2\omega}
\left(Z+1\right)^{-1/2+i2\wt{\omega}}
e^{-i  Y Z \frac{C_{BH}(\on)}{2} \delta N} \Biggr]\;.\nonumber \\
\label{Zints}
\eea

Let us first consider one of the $Z$ integrals (the top one),
\bea
&& \int\limits_{1}^\infty  dZ \left(Z+1\right)^{-1/2-i2\omega}
\left(Z-1\right)^{-1/2+i2\wt{\omega}}  e^{-i  Y Z \frac{C_{BH}(\on)}{2} \delta N}
\cr &=& \frac{1}{2} e^{-i  Y  \frac{C_{BH}(\on)}{2} \delta N} \int\limits_{1}^\infty  dZ \left(Z+1\right)^{-1/2-i2\omega}
\left(Z-1\right)^{-1/2+i2\wt{\omega}} e^{-i  Y (Z-1) \frac{C_{BH}(\on)}{2} \delta N}
\cr &+&
\frac{1}{2} e^{i  Y  \frac{C_{BH}(\on)}{2} \delta N} \int\limits_{1}^\infty  dZ \left(Z+1\right)^{-1/2-i2\omega}
\left(Z-1\right)^{-1/2+i2\wt{\omega}} e^{-i  Y (Z+1) \frac{C_{BH}(\on)}{2} \delta N}
\cr &=& \frac{1}{2}\left(e^{i  Y  \frac{C_{BH}(\on)}{2} \delta N}+ e^{-i  Y  \frac{C_{BH}(\on)}{2} \delta N}\right) f\left(\omega,\wt\omega;\delta N C_{BH}(\on)\right)\;,
\eea
where $f$ is a function that can be expressed in terms of gamma functions and hypergeometric functions.

However, the leading behavior
of the $Z$ integrals can be expressed in a simple way using the following considerations: The  $Z$ integrands are singular at $\;Z=\pm 1\;$
and well defined elsewhere, so we can expect that the main contribution to the integral comes from the vicinity of $Z=\pm 1$. These two contributions are equal in strength and are related by an exchange symmetry ($\omega\to-\widetilde{\omega}\;$, $\widetilde{\omega}\to-\omega$) that leaves the  $Y$ integral intact. The two $Z$ integrals are also equivalent up to the same exchange symmetry ($\omega\to-\widetilde{\omega}\;$, $\widetilde{\omega}\to-\omega$), which again leaves the  $Y$ integral  intact. Therefore, we can expect after integrating to pick up an extra overall  factor of
$\frac{1}{2}\left(e^{i  Y  \frac{C_{BH}(\on)}{2} \delta N}+ e^{-i  Y  \frac{C_{BH}(\on)}{2} \delta N}\right)$ plus subdominant corrections.

To understand the parameter that controls the strength of the corrections,
let us consider only  contributions  close to $Z=1$ (Similar arguments apply to $Z=-1$.) Then the phase can be written as
\be
e^{-i  Y (Z-1)  \frac{C_{BH}(\on)}{2} \delta N}\;=\;1 -i  Y (Z-1)  \frac{C_{BH}(\on)}{2} \delta N +{\cal O}\left[ \left((Z-1)  \frac{C_{BH}(\on)}{2} \delta N\right)^2\right]\;.
\ee
So that anything besides the leading-order result is suppressed by powers of  $\frac{C_{BH}(\on)}{2} \delta N$, as well as by powers of $Z-1$ which make the integral less singular.

And so the conclusion is that, to leading order in small parameters,
the $Z$ integral
 picks up a  $Y$-dependent phase factor relative to the time-independent
calculation,
\bea
&&\int\limits_{1}^\infty  dZ \left(Z+1\right)^{-1/2-i2\omega}
\left(Z-1\right)^{-1/2+i2\wt{\omega}} e^{-i  Y Z \frac{C_{BH}(\on)}{2} \delta N}
\nonumber \\  &+& \int\limits_{1}^\infty  dZ \left(Z-1\right)^{-1/2-i2\omega}
\left(Z+1\right)^{-1/2+i2\wt{\omega}} e^{-i  Y Z \frac{C_{BH}(\on)}{2} \delta N}
 \cr &=& \frac{1}{2}\left(e^{i  Y  \frac{C_{BH}(\on)}{2} \delta N}+ e^{-i  Y  \frac{C_{BH}(\on)}{2} \delta N}\right) \times \\ &&
\Biggl[\int\limits_{1}^\infty  dZ \left(Z+1\right)^{-1/2-i2\omega}
\left(Z-1\right)^{-1/2+i2\wt{\omega}}
+ \int\limits_{1}^\infty  dZ \left(Z-1\right)^{-1/2-i2\omega}
\left(Z+1\right)^{-1/2+i2\wt{\omega}}\Biggr]\;.\nonumber
\eea

The remaining integration over $Y$ in  Eq.~(\ref{Zints}) picks up an additional phase,
\bea
{\cal I}_Y(\on,\delta N)&=&\int\limits_{0}^\infty d Y \ e^{-\frac{Y^2}{4}C_{BH}(\on)}\
Y^{-i2(\omega-\wt{\omega})}\ e^{- i  Y  \frac{C_{BH}(\on)}{2}(N_T-\on)}\  \cr &\times& \frac{1}{2}\left(e^{i  Y  \frac{C_{BH}(\on)}{2} \delta N}+ e^{-i  Y  \frac{C_{BH}(\on)}{2} \delta N}\right)\;.
\eea
This integral can be expressed as a product of a gamma functions and a
 confluent hypergeometric function $U\left(\frac{1}{2}-i(\omega-\wt\omega), \frac{1}{2},-\frac{(C_{BH}(\on))^2\delta N^2}{4 C_{BH}(N_T)}\right)$.
However, its leading-order behavior can be determined by the following
argument:

By shifting the  integration variables to account for the phase factor
\be
e^{- i  Y  \frac{C_{BH}(\on)}{2}(N_T-\on)} \left(e^{i  Y  \frac{C_{BH}(\on)}{2} \delta N}+ e^{-i  Y  \frac{C_{BH}(\on)}{2} \delta N}\right)\;,\ee
one  finds a Gaussian times an exponentially decaying factor,
\be
{\cal I}_Y
\;=\; \frac{1}{2} \left(e^{-\frac{1}{4}\frac{\left[C_{BH}(\on) (N_T-\on-\delta N) \right]^2}{ C_{BH}(N_T)}}+ e^{-\frac{1}{4}\frac{\left[C_{BH}(\on) (N_T-\on+\delta N) \right]^2}{ C_{BH}(N_T)}}\right)\int\limits_{0}^\infty d Y \ e^{-\frac{Y^2}{4}C_{BH}(N_T)}\
Y^{-i2(\omega-\wt{\omega})}
\ee
plus subleading terms.

Hence, at leading order, the time-dependent density matrix is equal
to the time-independent matrix  of
Eq.~(\ref{rhoscf1}) multiplied by  the suppression factor
\bea
D(N_T,N^{\prime},N^{\prime\prime})&=&\frac{1}{2}\left(e^{-\frac{1}{4}\frac{\left[C_{BH}(\on) (N_T-\on-\delta N)\right]^2}{ C_{BH}(N_T)}}+e^{-\frac{1}{4}\frac{\left[C_{BH}(\on) (N_T-\on+\delta N)\right]^2}{ C_{BH}(N_T)}}\right) \cr & =&\;
\frac{1}{2}\left(e^{-\frac{1}{4} \frac{\left[C_{BH}(N^{\prime\prime}) (N_T-N^{\prime\prime})\right]^2}{ C_{BH}(N_T)}}+e^{-\frac{1}{4}\frac{\left[C_{BH}(N^{\prime}) (N_T-N^{\prime})\right]^2}{ C_{BH}(N_T)}}\right) \;.
\label{finalsf}
\eea


\begin{thebibliography}{99}

\bibitem{Bek}
 J.~D.~Bekenstein,
  ``Black holes and the second law,''
  Lett.\ Nuovo Cim.\  {\bf 4}, 737 (1972);
 ``Black holes and entropy,''
  Phys.\ Rev.\ D {\bf 7}, 2333 (1973);
  ``Generalized second law of thermodynamics in black hole physics,''
  Phys.\ Rev.\ D {\bf 9}, 3292 (1974).



\bibitem{Hawk} S. W. Hawking, ``Black hole explosions'',  Nature {\bf 248}, 30  (1974); ``Particle creation
by black holes'',
Comm. Math. Phys. {\bf 43}, 199 (1975).




\bibitem{info}
  S.~W.~Hawking,
  ``Breakdown of Predictability in Gravitational Collapse,''
  Phys.\ Rev.\ D {\bf 14}, 2460 (1976).

\bibitem{info2}
  D.~N.~Page,
  ``Black hole information,''
  hep-th/9305040.


\bibitem{info3}
  S.~B.~Giddings,
  ``Comments on information loss and remnants,''
  Phys.\ Rev.\ D {\bf 49}, 4078 (1994)
  [hep-th/9310101].



\bibitem{info4}
  S.~D.~Mathur,
  ``What Exactly is the Information Paradox?,''
  Lect.\ Notes Phys.\  {\bf 769}, 3 (2009)
  [arXiv:0803.2030 [hep-th]];
``The Information paradox: A Pedagogical introduction,''
  Class.\ Quant.\ Grav.\  {\bf 26}, 224001 (2009)
  [arXiv:0909.1038 [hep-th]];
 ``What the information paradox is {\it not},''
  arXiv:1108.0302 [hep-th].





\bibitem{page}
  D.~N.~Page,
  ``Average entropy of a subsystem,''
  Phys.\ Rev.\ Lett.\  {\bf 71}, 1291 (1993)
  [gr-qc/9305007];
 ``Information in black hole radiation,''
  Phys.\ Rev.\ Lett.\  {\bf 71}, 3743 (1993)
  [hep-th/9306083].

\bibitem{HaydenPreskill}
 P.~Hayden and J.~Preskill,
  ``Black holes as mirrors: Quantum information in random subsystems,''
  JHEP {\bf 0709}, 120 (2007)
  [arXiv:0708.4025 [hep-th]].




\bibitem{RB}
  R.~Brustein,
  ``Origin of the blackhole information paradox,''
  arXiv:1209.2686 [hep-th].
















\bibitem{Dvali1}
  G.~Dvali and C.~Gomez,
  ``Black Hole's Quantum N-Portrait,''
  arXiv:1112.3359 [hep-th];
  ``Black Hole's 1/N Hair,''
  Phys.\ Lett.\ B {\bf 719}, 419 (2013)
  [arXiv:1203.6575 [hep-th]].
  ``Black Holes as Critical Point of Quantum Phase Transition,''
   arXiv:1207.4059 [hep-th];
  ``Black Hole Macro-Quantumness,''
  arXiv:1212.0765 [hep-th].


\bibitem{Dvali2}
  G.~Dvali, C.~Gomez and D.~Lust,
  ``Black Hole Quantum Mechanics in the Presence of Species,''
  arXiv:1206.2365 [hep-th].




\bibitem{Dvali3}
 G.~Dvali, D.~Flassig, C.~Gomez, A.~Pritzel and N.~Wintergerst,
  ``Scrambling in the Black Hole Portrait,''
  arXiv:1307.3458 [hep-th].








\bibitem{flucyou}
  R.~Brustein and A.~J.~M.~Medved,
  ``Semiclassical black holes expose forbidden charges and censor divergent densities,''
  arXiv:1302.6086 [hep-th].





\bibitem{RM}  R.~Brustein and M.~Hadad,
  ``Wave function of the quantum black hole,''
  Phys.\ Lett.\ B {\bf 718}, 653 (2012)
  [arXiv:1202.5273 [hep-th]].





\bibitem{slowleak}
  R.~Brustein and A.~J.~M.~Medved,
  ``Restoring predictability in semiclassical gravitational collapse,''
  arXiv:1305.3139 [hep-th].









\bibitem{AMPS}
 A.~Almheiri, D.~Marolf, J.~Polchinski and J.~Sully,
  ``Black Holes: Complementarity or Firewalls?,''
  JHEP {\bf 1302}, 062 (2013)
  [arXiv:1207.3123 [hep-th]].

\bibitem{Sunny1}
 N.~Itzhaki,
  ``Is the black hole complementarity principle really necessary?,''
  hep-th/9607028.


\bibitem{Braun}
 S.~L.~Braunstein, S.~Pirandola and K.~Życzkowski,
  ``Entangled black holes as ciphers of hidden information,''
  Physical Review Letters 110, {\bf 101301} (2013)
  [arXiv:0907.1190 [quant-ph]].





\bibitem{fw1}
L.~Susskind,
  ``Singularities, Firewalls, and Complementarity,''
  arXiv:1208.3445 [hep-th];
  ``The Transfer of Entanglement: The Case for Firewalls,''
  arXiv:1210.2098 [hep-th].

 \bibitem{fw2}
  R.~Bousso,
  ``Complementarity Is Not Enough,''
  arXiv:1207.5192 [hep-th].




\bibitem{fw3}
Y.~Nomura, J.~Varela and S.~J.~Weinberg,
  ``Complementarity Endures: No Firewall for an Infalling Observer,''
  JHEP {\bf 1303}, 059 (2013)
  [arXiv:1207.6626 [hep-th]];
  ``Black Holes, Information, and Hilbert Space for Quantum Gravity,''
  arXiv:1210.6348 [hep-th].



\bibitem{fw4}
  S.~D.~Mathur and D.~Turton,
  ``Comments on black holes I: The possibility of complementarity,''
  arXiv:1208.2005 [hep-th].

\bibitem{Sunny2}
 A.~Giveon and N.~Itzhaki,
  ``String Theory Versus Black Hole Complementarity,''
  JHEP {\bf 1212}, 094 (2012)
  [arXiv:1208.3930 [hep-th]].



\bibitem{avery}
  S.~G.~Avery, B.~D.~Chowdhury and A.~Puhm,
  ``Unitarity and fuzzball complementarity: 'Alice fuzzes but may not even know it!',''
  arXiv:1210.6996 [hep-th].

\bibitem{lowe}
  K.~Larjo, D.~A.~Lowe and L.~Thorlacius,
  ``Black holes without firewalls,''
  arXiv:1211.4620 [hep-th].


\bibitem{vv}  E.~Verlinde and H.~Verlinde,
  ``Black Hole Entanglement and Quantum Error Correction,''
  arXiv:1211.6913 [hep-th];
  ``Passing through the Firewall,''
  arXiv:1306.0515 [hep-th].\;
``Black Hole Information as Topological Qubits,''
  arXiv:1306.0516 [hep-th].

\bibitem{pap}
  K.~Papadodimas and S.~Raju,
  ``An Infalling Observer in AdS/CFT,''
  arXiv:1211.6767 [hep-th].


\bibitem{AMPSS}
 A.~Almheiri, D.~Marolf, J.~Polchinski, D.~Stanford and J.~Sully,
  ``An Apologia for Firewalls,''
  arXiv:1304.6483 [hep-th].





\bibitem{SMfw}
 J.~Maldacena and L.~Susskind,
  ``Cool horizons for entangled black holes,''
  arXiv:1306.0533 [hep-th].

\bibitem{pagefw}
  D.~N.~Page,
  ``Excluding Black Hole Firewalls with Extreme Cosmic Censorship,''
  arXiv:1306.0562 [hep-th].


\bibitem{VRfw}  M.~Van Raamsdonk,
  ``Evaporating Firewalls,''
  arXiv:1307.1796 [hep-th].

\bibitem{MP} D. Marolf and J. Polchinski,
``Gauge/Gravity Duality and the Black Hole Interior,''
arxiv:1307.4706 [hep-th].








\bibitem{inprog} R. Brustein and A. J. M. Medved, work in preparation.


\bibitem{future} R. Brustein and A. J. M. Medved, work in progress.




\bibitem{Peres}
  A.~Peres,
  ``Separability criterion for density matrices,''
  Phys.\ Rev.\ Lett.\  {\bf 77}, 1413 (1996)
  [quant-ph/9604005].

\bibitem{Horod}
  M.~Horodecki, P.~Horodecki and R.~Horodecki,
  ``On the necessary and sufficient conditions for separability of mixed quantum states,''
Phys. Lett. A {\bf 223}, 1 (1996)
[quant-ph/9605038].



\end{thebibliography}
\end{document}